\documentclass[twocolumn]{aastex63}
\usepackage{amsmath}
\newcommand{\be}{\begin{eqnarray}}
\newcommand{\ee}{\end{eqnarray}}
\def\la{\mathbin{\lower 3pt\hbox
      {$\rlap{\raise 5pt\hbox{$\char'074$}}\mathchar"7218$}}}
\def\ga{\mathbin{\lower 3pt\hbox
      {$\rlap{\raise 5pt\hbox{$\char'076$}}\mathchar"7218$}}}
\renewcommand{\vec}[1]{\ensuremath{\boldsymbol{#1}}} 
\newcommand{\grad}{\mathbf{\nabla}}

\defcitealias{Press:77}{PT77}
\defcitealias{Vick:18}{VL18}

\submitjournal{ApJ}

\shorttitle{}
\shortauthors{}

\begin{document}

\title{ Large Dynamical Tide Amplitudes from Small Kicks at Pericenter }

\correspondingauthor{Phil Arras}
\email{arras@virginia.edu}

\author[0000-0001-5611-1349]{Phil Arras}
\affiliation{Department of Astronomy, University of Virginia, Charlottesville, VA 22904, USA}

\author[0000-0002-6011-6190]{Hang Yu}
\affiliation{Kavli Institute for Theoretical Physics, University of California at Santa Barbara, Santa Barbara, CA 93106, USA}

\author[0000-0001-9194-2084]{Nevin N. Weinberg}
\affiliation{Department of Physics, University of Texas at Arlington, Arlington, TX 76019, USA}

\begin{abstract}
The effect of  dynamical tide ``kicks" on eccentric binary orbits is considered using the orbital mapping method. It is demonstrated that when mode damping is negligible the mode amplitude will generically grow in time for all values of orbital eccentricity and semi-major axis, even for small kicks outside the regime exhibiting diffusive growth. The origin of the small-kick growth is the  change in kick size from orbit to orbit, an effect quadratic in the mode amplitude. When damping of the mode is included, the growth is shut off when the damping time is shorter than the growth time. Hence, in practice, kicks of sufficient size and long mode damping times are required for interesting levels of growth to occur. Application to the circularization of hot Jupiters is discussed. Previous investigations found that diffusive growth of the planetary f-mode in the large-kick regime would lead to rapid orbital shrinkage, but upon exiting the diffusive regime at $e \sim 0.9$ the theory would predict a large population of highly eccentric orbits. Simulations presented here show that subsequent orbital evolution relying on the small-kick regime may further decrease the eccentricity to $e \sim 0.2$ on timescales much less than the Gyrs ages of these systems.
\end{abstract}

\keywords{}

\section{Introduction}
\label{sec:intro}

Many binaries begin their lives on highly eccentric orbits that circularize over time due to tides. The rate of circularization depends on the efficiency of tidal dissipation, an often complicated process that has been studied in a variety of systems.  These include tidal capture binaries (see, e.g., \citealt{Press:77, Kumar:96, Lai:97, Samsing:17}), solar-type binaries \citep{Goodman:98, Terquem:98, Terquem:21, Zanazzi:21, Barker:22}, white dwarf binaries \citep{Willems:07, Vick:17, McNeill:20, Lau:22}, neutron star binaries \citep{Kumar:95, Gold:12, Yang:18, Vick:19b}, and hot Jupiter systems \citep{Ivanov:04, Wu:18, Vick:19, Yu:21, Yu:22}.

For sufficiently eccentric orbits, the internal oscillation modes excited at pericenter exhibit a random walk in mode energy called diffusive growth \citep{Kochanek:92,Mardling:95}. \citet{Mardling:95} found that the coupled evolution of the orbit and oscillation modes exhibits chaos. This regime occurs when the tidal ``kick", which transfers energy from the orbit to the mode, is so large that it substantially alters the orbital period from orbit to orbit. The phase of the oscillation mode at each pericenter passage is then effectively random and the mode energy increases linearly in time as in a random walk. If nonlinear three-wave coupling effects are included, the energy absorbed by the mode alters its oscillation frequency due to the anharmonicity, which can also produce a random phase at each pericenter passage and trigger the diffusive growth~\citep{Yu:21}.  

The original study by \citeauthor{Mardling:95} (1995a; see also \citealt{Mardling:95b}) focused primarily on tidal capture binaries.  Since then studies have shown that this process
can also be important in neutron star binaries \citep{Vick:19b},  hierarchical triple systems \citep{Mardling:01, Hayashi:22},  gaseous planets orbiting white dwarfs \citep{Veras:19}, and in the high eccentricity migration of hot Jupiters (see last set of citations in previous paragraph).

For small kicks, neither the backreaction of the mode on the orbit nor the anharmonic shift of the mode's frequency is sufficient to trigger diffusive growth. 
Previous studies found that in this small kick regime, 
the mode undergoes periodic changes in amplitude and does not grow \citep{Mardling:95, Vick:18, Wu:18}. According to these studies, as the orbit circularizes and the kick decreases, the system leaves the chaotic regime and the tidal evolution slows dramatically. In the hot Jupiter context, \citeauthor{Wu:18} (2018; see also \citealt{Yu:22}) found that in the chaotic regime, the orbital eccentricity can evolve quickly (on timescales of $\sim 10^4\textrm{ year}$) from $e\gtrsim 0.98$ to  $e\simeq 0.9$ but then stalls at $e\simeq 0.9$.  In order to produce the circular orbits of observed hot Jupiters, a different, unidentified tidal circularization process must take over and  drive the orbit from $e\simeq 0.9$ to $e\simeq 0$.

When evaluating the small kick regime,  previous studies only considered timescales $\lesssim 10^4$ orbits or assumed a constant kick (e.g. \citealt{Vick:18,Wu:18}).  In this paper, we show that when the kick size is  allowed to vary from orbit to orbit, the mode's amplitude changes are not necessarily periodic.  Rather, they can grow linearly in time, typically on timescales $\gtrsim 10^4$ orbits.  In the absence of mode damping, this occurs for all
values of $e$ and semi-major axis.  When mode damping is included,  the amplitude can grow if the growth rate exceeds the damping rate. As we will show, this has potentially important implications for the continued tidal circularization of binaries after they leave the chaotic regime.

A detailed understanding of the change in an oscillation mode over an orbit is required to implement the mapping method. Most previous investigations have used the kick formula from \citeauthor{Press:77} (1977; hereafter \citetalias{Press:77}) derived for an unbound (parabolic) orbit. One exception was \citet{Wu:18}, who presented results for a bound (elliptic) orbit kick. However, due to the unexpected ``resonant" behavior found in those results, only the PT77 formula was used in the later analytic results. Of course, the numerical orbit integrations presented in that study self-consistently evaluated the pericenter passage for a bound orbit. Here we revisit the kick for a bound orbit from one apocenter to the next. It is shown that there are two physically distinct contributions to the change in mode amplitude over an orbit, that from the equilibrium tide at apocenter, and from the dynamical tide kick at pericenter. We show that the latter reduces to the \citetalias{Press:77} result in the limit  $e \rightarrow 1$. The former explains the ``resonance" features found in \citet{Wu:18}. The two constributions are evaluated separately using contour integration in the Appendix. 

The plan of the paper is as follows. Section \ref{sec:statement} reviews the dynamical system under consideration, the kick in mode amplitude each orbit, the backreaction of the modes on the orbit, and the physical origin of the small and large kick regimes. Section \ref{sec:growth} demonstrates the evolution of mode amplitude in the small and large kick regimes, as well as the crucial role of the variation in kick size from orbit to orbit. To explore the physical origin of growth in the small kick regime, Section \ref{sec:analytic} suggests a toy model with the main elements. Section \ref{sec:damping} shows, through an analytic derivation and simulations, that mode damping cannot suppress the growth if the damping time is longer than the growth time. Section \ref{sec:results} show results for a Jupiter-size planet orbiting a solar-like star. 
Section \ref{sec:summary} presents the summary and conclusions. The Appendix contains a detailed discussion of the one-kick integral.

\section{ statement of the problem }
\label{sec:statement}

\subsection{ Equations of motion }

Consider masses $M_1$ and $M_2$ with radii $R_1$ and $R_2$, and an orbit in the x-y plane with cylindrical coordinates $(r,\phi)$. Star 1 is perturbed by the tidal gravity of star 2. An oscillation mode labeled $a$ in star 1 has complex amplitude $q_a$ and gives rise to a physical displacement  $\vec{\xi}(\vec{x},t) = q_a(t)  \vec{\xi}_a(\vec{x}) + \rm c.c.$ from the unperturbed to perturbed star. The azimuthal index and frequency of this mode are $m_a$ and $\omega_a$. The set of coupled equations for the orbit and mode amplitude are \citep{Weinberg:12}
\be
\ddot{r} & = & \frac{l^2}{r^3} - \frac{GM}{r^2} + a_r, \label{eq:ddotr} \\
\dot{l} & = & r a_\phi, \label{eq:elldot} \\
\dot{\phi} & = & \frac{\ell}{r^2} \label{eq:phidot} \\
\dot{q}_a & = & - i \omega_a q_a - \gamma_a q_a + i \omega_a U_a, \label{eq:qdot}
\ee
where $M=M_1+M_2$ is the total mass, $l$ is the orbital angular momentum per unit mass, $a_r$ and $a_\phi$ are the radial and tangential accelerations of the orbit due to the perturbed gravity created by the oscillation mode, and $U_a(t)$ is the projection of the tidal force along mode $a$. 
The energy and angular momentum of the oscillation mode, including  its physically identical complex conjugate, are $(GM_1^2/R_1)|q_a |^2$ and $(GM_1^2/R_1)|q_a |^2(m_a/\omega_a)$. Excitation of the oscillation mode causes an exchange of energy and angular momentum with the orbit. Mode energy and angular momentum are damped into the background at rate $\gamma_a$. 

The tidal driving of mode $a$  is given by
\be
U_a &= & \frac{M_2}{M_1} W_{\ell m} I_{a \ell m} \left( \frac{R_1}{r} \right)^{\ell+1} e^{-i m \phi},
\ee
where the coefficient $W_{\ell m}=4\pi Y_{\ell m}(\pi/2,0)/(2\ell+1)$  
and the  integral
\citep{Weinberg:12}
\be
I_{a\ell m} &= & \frac{1}{M_1 R_1^\ell} \int d^3x \rho \vec{\xi}_a^* \cdot \grad \left( r^\ell Y_{\ell m} \right)
\ee
quantifies the overlap of  mode $a$ with the tidal force, or equivalently the size of the mode's multipole moments.
For the case considered here, where rotation is ignored, $I_{a \ell m} \propto \delta_{m,m_a} \delta_{\ell \ell_a}$.
The eigenfunctions are normalized to energy $GM_1^2/R_1$ at unit amplitude $|q_a|=1$.
The accelerations acting on the orbit due to mode $a$ are
\be
a_r & = & - \frac{GM}{R_1^2}(\ell+1)W_{\ell m}I_{a\ell m}q_a^* \left( \frac{R_1}{r} \right)^{\ell+2} e^{-i m \phi}
+ c.c.
\\
a_\phi & = & - \frac{GM}{R_1^2}(im)W_{\ell m}I_{a\ell m}q_a^* \left( \frac{R_1}{r} \right)^{\ell+2} e^{-i m \phi}
+ c.c.
\ee
The system of equations can be written as 6 first order ordinary differential equations for the variables $r$, $\dot{r}$, $\phi$, $l$, ${\rm Re}(q_a)$ and ${\rm Im}(q_a)$. Additional mode amplitude equations can be added, with their effects included in $a_r$ and $a_\phi$. Modes in star 2 can be included through interchange of the labels 1 and 2, as well as letting $\phi \rightarrow \phi + \pi$. 

\subsection{ Approximations }
\label{sec:approximations}

In the absence of wave-background interaction, i.e. ignoring wave damping and corotation resonances \citep{Goldreich:77}, energy and angular momentum are conserved for the system of modes and orbit (e.g. \citealt{Mardling:95}). For angular momentum, the sum of mode angular momentum and orbital angular momentum is constant, while for energy it is the sum of mode energy, orbital energy, and interaction energy that is constant. In previous papers using the mapping method, the interaction energy has typically been ignored, which is formally only valid for an unbound orbit where bodies proceed to infinite separation. For a bound orbit, the interaction energy term should in principle be included, and may change from one apocenter passage to the next. 
We also ignore the effect of changes in the other orbital elements such as the longitude of the pericenter or the epoch of periastron. In principle, these quantities could change due to the dynamical tide at successive pericenter passages, and hence could contribute to the changes in mode phase needed for diffusive growth. 
We continue to ignore these effects in the current study, but plan to include them in future work \citep{vincent2023}. Lastly, we ignored the changes in the background planet due to wave-background interaction, including both energy and angular momentum exchange (e.g. \citealt{Wu:18}). These deserve attention in a future study.

\subsection{ Changes in the orbit due to mode growth }

Given these approximations, the changes in semi-major axis $a$ (not to be confused with the mode index) and eccentricity $e$ may then be computed as a function of mode amplitude $q_a$, giving simple expressions in the limit of small mode amplitude. 
Let $a_0$ and $e_0$ be the semi-major axis and eccentricity of the orbit when the mode has zero energy $q_a=0$, while $a$ and $e$ are the values for finite $q_a$. It is convenient to define a rescaled amplitude $Q_a = \sqrt{2(M_1/M_2)(a_0/R_1)}q_a$, for which $|Q_a|^2=1$ implies the mode energy equals the orbital energy $GM_1M_2/2a_0$. We will use $q_a$ and $Q_a$ interchangeably, and will drop the mode index $a$ when it does not cause confusion.
The conservation laws then imply
\be
a &= & a_0  \left( 1 + |Q_a|^2 \right)^{-1}, \label{eq:a_of_Q}\\
1 & = & \left( \frac{a(1-e^2)}{a_0(1-e_0^2)} \right)^{1/2}
+ \frac{m_a n_0}{2\omega_a} |Q_a|^2 \left( 1 - e_0^2 \right)^{-1/2} \label{eq:e_of_Q}.
\ee
Here $n_0 =  \sqrt{GM/a_0^3}$ and $P_0=2\pi/n_0$ are the mean motion and period of the orbit at zero mode amplitude.
For the demonstration of mode growth at small amplitude, it is convenient to work 
in the limit of small $|Q_a|^2$, for which the small deviations $\delta a = a-a_0$ and $\delta e = e-e_0$ simplify to
\be
\frac{\delta a}{a_0} &\simeq & - |Q_a|^2 
\label{eq:da} \\
\delta e & \simeq & \left( \frac{|Q_a|^2}{2e_0} \right) \left[ - (1-e_0^2) + \frac{m_a n_0}{\omega_a} \sqrt{1-e_0^2} \right].
\label{eq:de}
\ee
The associated change in orbital period is then $\delta P/P_0 = (3/2) \delta a/a_0$,
while the pericenter shift is
\be
\frac{ \delta r_p}{r_{p,0}} & \simeq &  \left( \frac{|Q_a|^2}{2e_0} \right) \left[ \left( 1-e_0 \right) - \frac{m_a n_0}{\omega_a} \sqrt{\frac{1+e_0}{1-e_0}} \right].
\ee
In $\delta r_p$, the two terms may be comparable implying that the angular momentum input to the mode, and taken out of the orbit, should not be neglected.

\subsection{ The ``exact" mapping method }

The tidal forces on the mode are largest at pericenter and the mode receives a kick $\Delta q_a$ during the pericenter passage. We define the ``exact" orbital mapping technique to consist of the following three steps \citep{Vick:18}:
\be
&& {\rm (i)\, Let\, } q_a {\rm \, represent\, the\, mode\, amplitude\, just\, before\, }
\nonumber \\ && {\rm pericenter.} \label{eq:step1} \\
&& {\rm (ii)\, }\, q_a+\Delta q_a {\rm\, is\, the\, amplitude\, just\, after\, pericenter\, }
\nonumber \\ && {\rm and\, determines\, the\, orbital\, period\,} P {\rm \, and\, eccentricity}
\nonumber \\ &&  e {\rm \, for\, the\, ensuing\, orbit.} \label{eq:step2} \\
&& {\rm (iii)\, The\, mode\, undergoes\, free\, oscillation\, over\, the\, rest\, }
\nonumber \\ && {\rm of\, the\, orbit,\, and\, the\, amplitude\, just\, before\, the\, next\,}
\nonumber \\ && {\rm pericenter\, passage\, is\,} (q_a+\Delta q_a)\exp(-i \omega_a P ). \label{eq:step3}
\ee
Damping may be included in step (iii) through a factor $\exp(-\gamma_a P)$.
For simplicity, in this paper we ignore the anharmonic frequency shift of the mode though formally it comes in at the same order as the tidal backreaction as $\delta \omega_a /\omega_a \propto |Q_a|^2$. In step (ii), conservation of orbital plus mode energy, and conservation of orbital plus mode angular momentum, are used to related the pre- and post-pericenter passage values of $a$ and $e$. 

\subsection{ Kicks at pericenter}

\begin{figure}[htb]
    \epsscale{1.1}
    \plotone{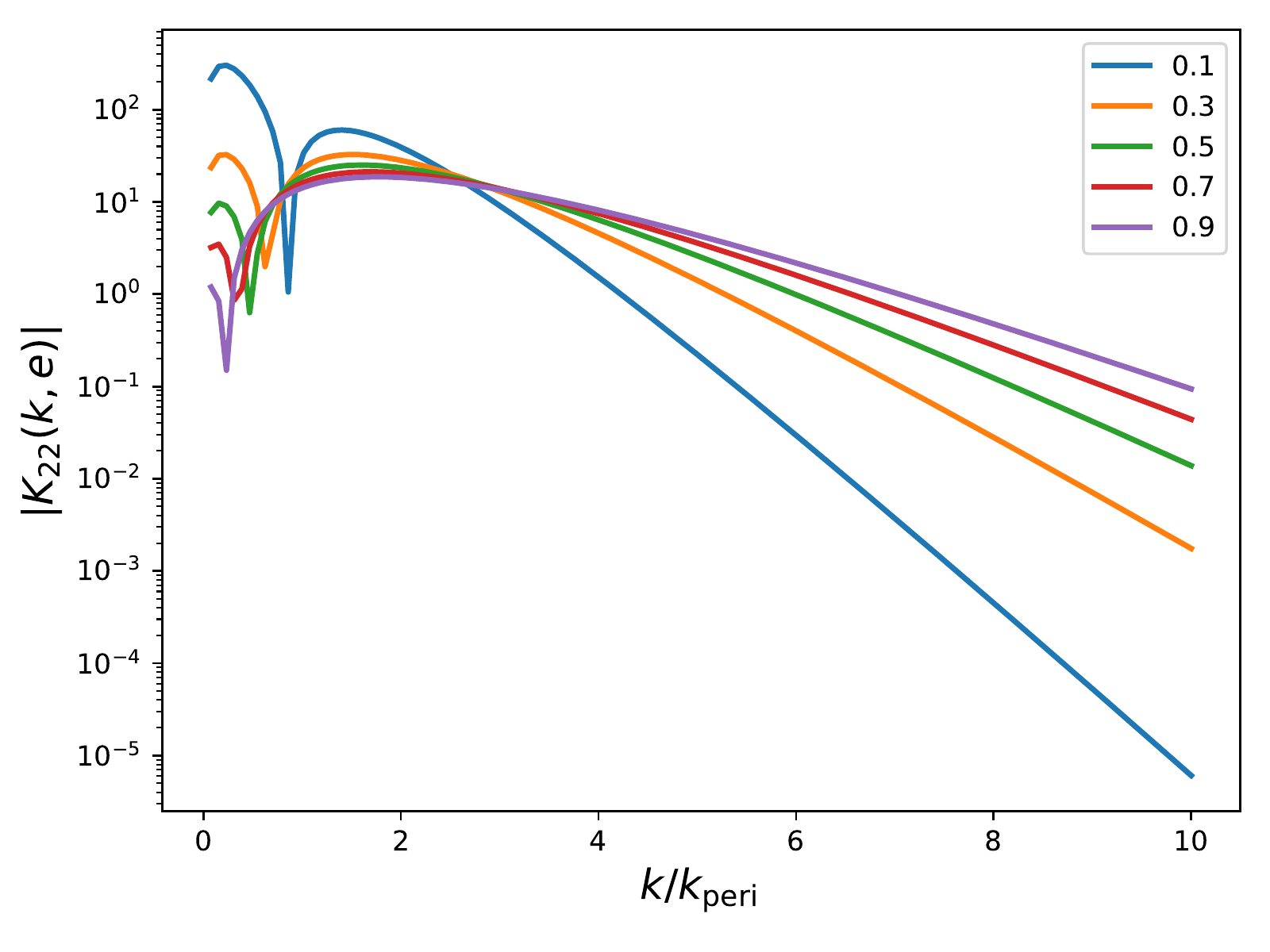} 
    \caption{Dynamical tide piece of the one-kick amplitude for $\ell=m=2$ as a function of $k$. The different lines are labeled by the eccentricity.  }
    \label{fig:onekick_e_comparison}
\end{figure}

Equation \ref{eq:qdot} may be integrated to give the kick $\Delta q_a$ for a bound Keplerian orbit and eccentricity $e$, with the result
\be
\Delta q_a & = & i \frac{M_2}{M_1} W_{\ell m} I_{a \ell m} \left( \frac{R_1}{r_p} \right)^{\ell+1} K_{\ell m}\left(\frac{\omega_a}{n},e\right)
\label{eq:kick}
\ee
where the function
\be
K_{\ell m}(k,e) & = & k (1-e) \int_{-\pi}^\pi \, du\, \left( \frac{ 1-e}{1-e\cos u} \right)^\ell \nonumber \\ 
& \times & 
\cos \left[ k (u - e \sin u) - m f(u) \right].
\label{eq:Klm}
\ee
Here $n=\sqrt{GM/a^3}$ is the mean motion, $r_p=a(1-e)$ is the pericenter distance, $u$ is the eccentric anomaly, $u-e\sin u$ is the mean anomaly, and $f(u)$ is the true anomaly. For integer $k=\omega_a/n$, $K_{\ell m}(k,e)$ is related to the well-known Hansen coefficients (e.g. \citealt{2000ssd..book.....M}) which have exponentially small values for $|k| \gg 1$. This is the dynamical tide component, and it arises from forcing of the mode near pericenter. In the Appendix, a detailed discussion of Equation \ref{eq:Klm} is given, including both the equilibrium and dynamical tide contributions, as well as an efficient numerical method for their calculation. 

Figure \ref{fig:onekick_e_comparison} shows the dynamical tide contribution to $K_{22}(k,e)$ as function of $k/k_{\rm peri}$ for a range of eccentricity. Here $k_{\rm peri} \equiv (1-e)^{-3/2}$ is the harmonic number corresponding to the pericenter passage. The dynamical tide kick has roughly the same peak value for different $e$, but decreases faster at large $k$ for small $e$.  The parabolic  approximation  may overestimate the kicks if applied to an orbit with $e \la 0.9$. In the Appendix, it is shown that for an elliptic ($0<e<1$)  orbit and $|k| \gg 1$, the dominant dependence in the exponent is given by
\be
K_{\ell m}(k,e) & \propto & e^{-k(\cosh^{-1}(1/e) - \sqrt{1-e^2})} \label{eq:Klm_exponent}
\ee
with a prefactor that involves a polynomial in $k^{1/2}$ and functions of eccentricity. The exponent reduces to $-(2^{3/2}/3)k(1-e)^{3/2}$ as $e \rightarrow 1$, as found by \citetalias{Press:77}. At small $e$, the exponential
factor recovers the power-law $e^k$ behavior expected at small eccentricity.

For non-integer $k=\omega_a/n$, as is the case here when no resonance occurs, $K_{\ell m}(k,e)$ will still have the dynamical tide component, but a physically distinct piece arises, as first found in \citet{Wu:18}, which can be identified as coming from the equilibrium tide at apocenter. That piece involves a factor $\sin(\pi k)$, and hence is zero for integer $k$, and exhibits the resonant feature found in \citet{Wu:18}. For larger pericenter distance, or smaller eccentricities, the dynamical tide piece can be dwarfed by the equilibrium tide piece, but it is still present. 

As the equilibrium tide is strictly a function of the instantaneous orbital position, it is not an additional degree of freedom. It provides an additional radial force acting on the orbit that changes the relationship between orbital period and radius away from the Keplerian value. When dissipative effects are included, the equilibrium tide may be an important driver of orbital evolution (e.g. \citealt{1989A&A...220..112Z,Ivanov:04}). However, in the inviscid limit considered here it has no effect on the secular evolution of the orbit. The dynamical tide, by contrast, arises due to the additional degrees of freedom from the mode amplitude, and it can cause long-term changes in the orbit as energy and angular momentum are shared between orbital and mode degrees of freedom. Henceforth, the equilibrium tide kick will be ignored and the mapping will use only the dynamical tide kick.

\subsection{ Critical kick and amplitude }

The effect of dynamical tides is a strongly decreasing function of the ratio of pericenter crossing time to mode period, and large kicks require such close encounters that the body is nearly filling its Roche lobe with $R_1 \sim r_p (M_1/M_2)^{1/3}$. Starting from small amplitude, the diffusive limit in which the variation of the mode phase  is large from orbit to orbit, requires \citep{Vick:18}
\be
\omega_a \delta P & \sim \omega_a P_0 |\Delta Q_a|^2 \ga 1
\ee
and hence a  critical kick size separating the large and small kick limits is 
\be
   \Delta Q_{\rm a, cr} & \equiv &  \left( \omega_a P_0 \right)^{-1/2}
  \nonumber \\ 
&  \simeq & 10^{-2}\, \left( \frac{ \sqrt{GM_{\rm J}/R_{\rm J}^3} }{\omega_a} \right)^{1/2} \left( \frac{1\, \rm yr}{P_0} \right)^{1/2},
 \ee
 where the fiducial value uses the dynamical frequency $\omega_a \simeq (GM_{\rm J}/R_{\rm J}^3)^{1/2}$ of Jupiter and a 1 year orbital period. Short period orbits have large $\Delta Q_{\rm a, cr}$, which disfavors the diffusive limit, and vice versa for long-period orbits. We caution that $\Delta Q_{\rm a, cr}$ is only a rough criterion to separate the two limits.

\citet{Wu:18} has shown that even if $\Delta Q_{\rm a} < \Delta Q_{\rm a, cr}$, diffusion can still occur if the mode amplitude becomes larger than a critical value
\be
Q_{\rm a, cr} & \equiv & \frac{\Delta Q_{\rm a, cr}^2}{\Delta Q_{\rm a}},
\ee
which is larger than $\Delta Q_{\rm a, cr}$. This latter criterion is relevant to the small kick limit in which $\Delta Q_{\rm a} < \Delta Q_{\rm a, cr}$, as the small kick growth can lead to sufficiently large amplitudes that diffusion may begin, after which the growth is faster. Alternatively, if $q_{\rm a, cr} $ is larger than the wave-breaking amplitude $q_{\rm max}$, small kick growth may proceed all the way to nonlinear amplitudes without diffusive behaivor.

\section{ mode growth in the large and small kick regimes }
\label{sec:growth}

In this section, examples will be shown of mode growth in the large and small kick limits, ignoring the effects of damping. 

\subsection{ Outline of the two regimes }

A brief summary of the results is as follows. When $\Delta Q \ga \Delta Q_{\rm cr}$, the mode phase is effectively random between pericenter passages and a random walk in mode energy energy occurs \citep{Mardling:95, Vick:18, Wu:18}
\be
|Q_j^{(\rm diff)}| & \sim & |\Delta Q|\, j^{1/2},
\ee
where $j$ is the orbit number. This random walk in mode energy is called diffusive or chaotic growth. In this regime, growth of mode energy to the size of the orbital energy is fast and requires only $j \sim \Delta Q^{-2}$ orbits. By contrast, when $\Delta Q \la \Delta Q_{\rm cr}$, the change in mode phase between pericenter passages is much smaller than unity, and diffusive growth does not occur. Previous investigations found that on short timescales the mode exhibited strictly periodic changes in amplitude from one orbit to the next. However, it will be shown in Section \ref{sec:analytic} that, generically, linear growth in the mode amplitude
\be
|Q_j^{(\rm small)}| & \sim & |\Delta Q| ^3 j
\ee
occurs in this small kick regime. The time to grow to large amplitudes is then $j \sim \Delta Q^{-3}$, which is longer both due to the higher power of $\Delta Q$ but also because $\Delta Q \la \Delta Q_{\rm cr}$. Nevertheless, as will be shown in this section, the small kick instability can still operate and cause growth to large amplitudes.

\subsection{ Numerical examples }

\begin{figure}[htb]
    \epsscale{1.1}
    \plotone{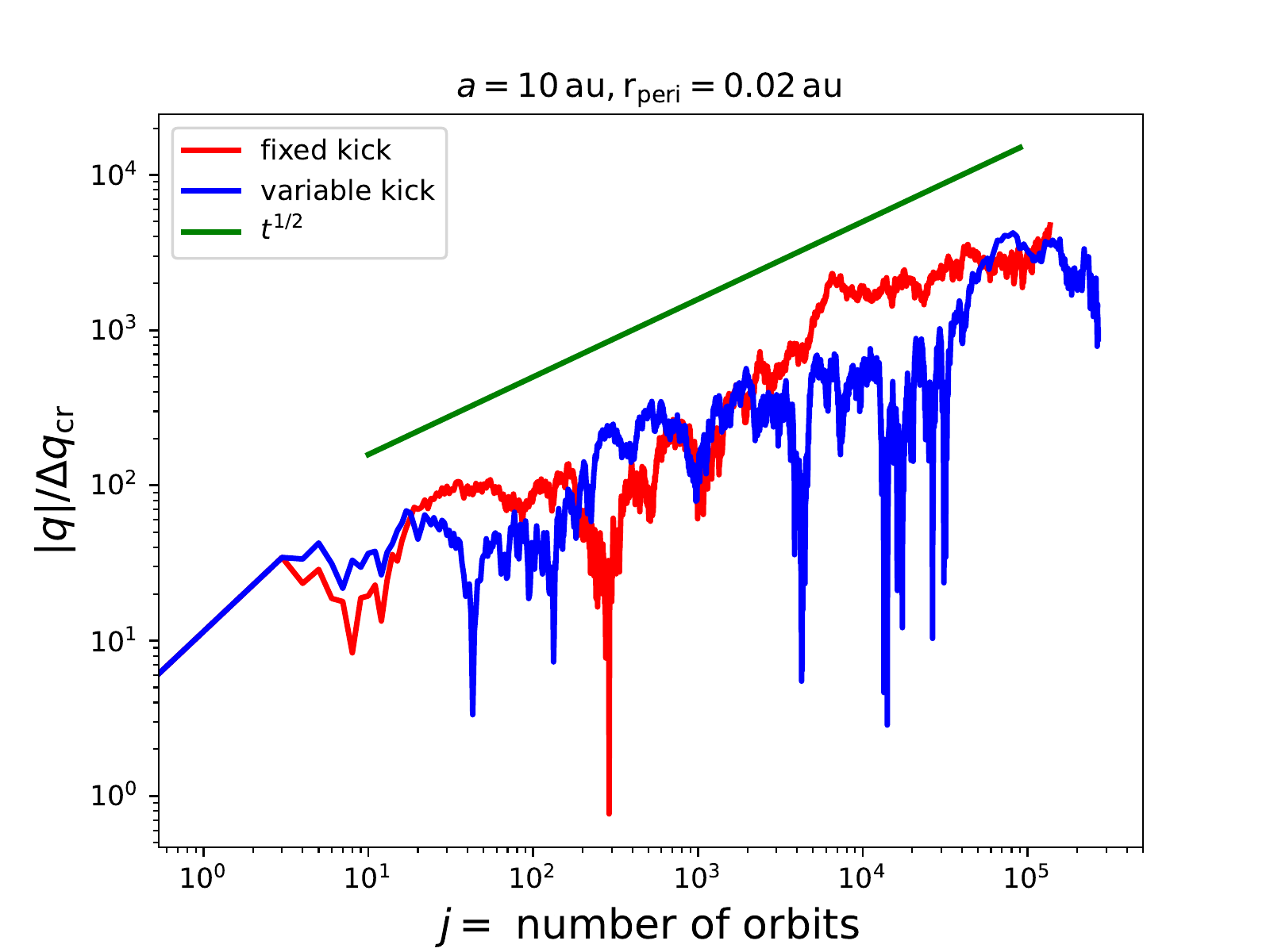} 
    \caption{Mode amplitude versus orbit number for $a_0=10\, \rm AU$ and $r_{p,0}=a_0(1-e_0)=0.02\, \rm AU$. For the red line, the kick is held fixed to that of the initial orbit. For the blue line, the kick is allowed to vary as the orbit varies, and the mode angular momentum is included. The green line shows the expected growth in the diffusive limit. The noisy appearance of each line is characteristic of the diffusive regime in which the mode energy undergoes a random walk.  }
    \label{fig:q_vs_t_10AU_0.02AU}
\end{figure}

\begin{figure*}[ht]
    \epsscale{1.1}
    \plotone{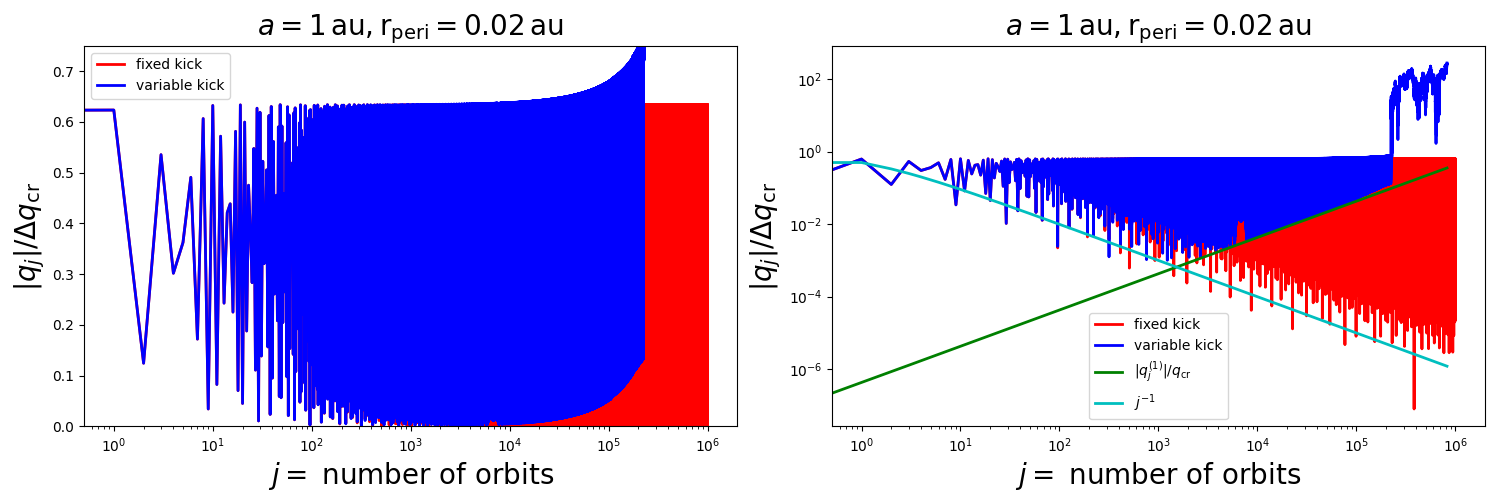} 
    \caption{(Left panel) Same log-linear plot as Figure \ref{fig:q_vs_t_10AU_0.02AU}, but for $a_0=1\, \rm AU$. In this case the change in mode phase is insufficient to cause diffusive growth of the mode as seen in Figure \ref{fig:q_vs_t_10AU_0.02AU}. (Right panel) Log-log axes where the blue line is the same as in the left panel, and the green line is the analytic prediction from Equation \ref{eq:Qk1}.  } 
    \label{fig:q_vs_t_1AU_0.02AU}
\end{figure*}

In this section, numerical calculations will be presented using the ``exact" mapping method (Equations \ref{eq:step1}-\ref{eq:step3}) and the dynamical tide kick for a Keplerian orbit (Equations \ref{eq:kick} and \ref{eq:Klm}). The angular momentum of the mode is included, so that the orbital angular momentum varies.  The parameters used are $M_1=10^{-3}\ M_\odot$ and $R_1=10^{-1}\, R_\odot$ for the planet, and $M_2=M_\odot$ 
for the star. For Roche-lobe radius $R_{\rm rl, 1} \simeq 0.49a(M_1/M_2)^{1/3}$ \citep{1983ApJ...268..368E}, the planet would fill its Roche lobe at a separation $a=2\, R_\odot=10^{-2}\, \rm AU$. A single oscillation mode representing the $\ell=m=2$, prograde f-mode is included in the the planet (body 1) with mode frequency 
$\omega_a = \sqrt{GM_1/R_1^3}$ and overlap integral $I_{a \ell m}=0.4$ \citep{Yu:21}. The simulations are stopped either when $|q_a|=1$ or a maximum run time is reached.

Figure \ref{fig:q_vs_t_10AU_0.02AU} shows mode amplitude versus orbit number ($j \propto t$) for the case of strong kicks well above the critical value. The amplitude is observed to undergo irregular changes and increases roughly with the slope $|q_a| \propto j^{1/2}$. The initial orbit has $a_0=10\, \rm AU$ and 
$e_0=0.998$, with initial pericenter $a_0(1-e_0)=0.02\, \rm AU$. The red line uses constant kick given by the initial orbit while the blue line allows the kick to vary as the orbital energy and angular momentum vary. Initially the two curves are in agreement, but they later diverge, as expected of chaotic orbits receiving slightly different kick values. Both show the expected linear growth of energy with time.

Figure \ref{fig:q_vs_t_1AU_0.02AU} uses a smaller $a_0=1\, \rm AU$ and $e_0=0.98$, with the same initial pericenter $a_0(1-e_0)=0.02\, \rm AU$. The initial kick amplitude is roughly the same, but the orbital energy is 10 times larger, so that diffusive growth does not occur at small mode amplitude. The kick amplitude is near, but smaller than the critical value. The red line (fixed kick) shows oscillations in the amplitude for $10^6$ orbits with no sign of growth. The blue line (variable kick) initially also appears to show oscillations in the amplitude, agreeing with the red line. However, after $\sim 10^4$ orbits a trend of increasing amplitude is apparent, and eventually a rapid runaway due to diffusive growth occurs near $2\times 10^5$ orbits. Hence this ``periodic" regime is not truly periodic, as the amplitude still grows, albeit more slowly than seen in Figure \ref{fig:q_vs_t_10AU_0.02AU}. The key difference allowing mode growth is the variation of the kick as the orbit varies. 

The left panel of Figure \ref{fig:q_vs_t_1AU_0.02AU} uses log-linear axes to more clearly see the upper envelope of the mode amplitude. It is clear that the red curve (fixed kick) is not growing in time, and we have verified this over even longer run times not shown here. The right panel of Figure \ref{fig:q_vs_t_1AU_0.02AU} uses log-log axes to more clearly see small mode amplitudes. The decreasing lower envelope of the red curve (fixed kick) is essentially due to sampling a sine curve at random angles (see Equation \ref{eq:oscillatory}). The more trials (larger $j$) the closer the points get on average to $0$ or $\pi$, and the smaller the value of the sine. This gives the $|q_j| \propto 1/j$ envelope plotted for comparison (cyan line). The blue line (variable kick) initially shows the $q_j \propto 1/j$ decrease, but eventually the envelope begins to increase as $q_j \propto j$. In Section \ref{sec:analytic}, an analytic model is developed which shows this linear growth, with Equation \ref{eq:Qk1} showing good agreement with the lower envelope observed in the numerical results (green curve). Lastly, note that diffusive growth does not occur in the plot as the amplitude has not passed the critical amplitude $q_{\rm a, cr}$.

\begin{figure*}[htb]
    \epsscale{1.1}
    \plotone{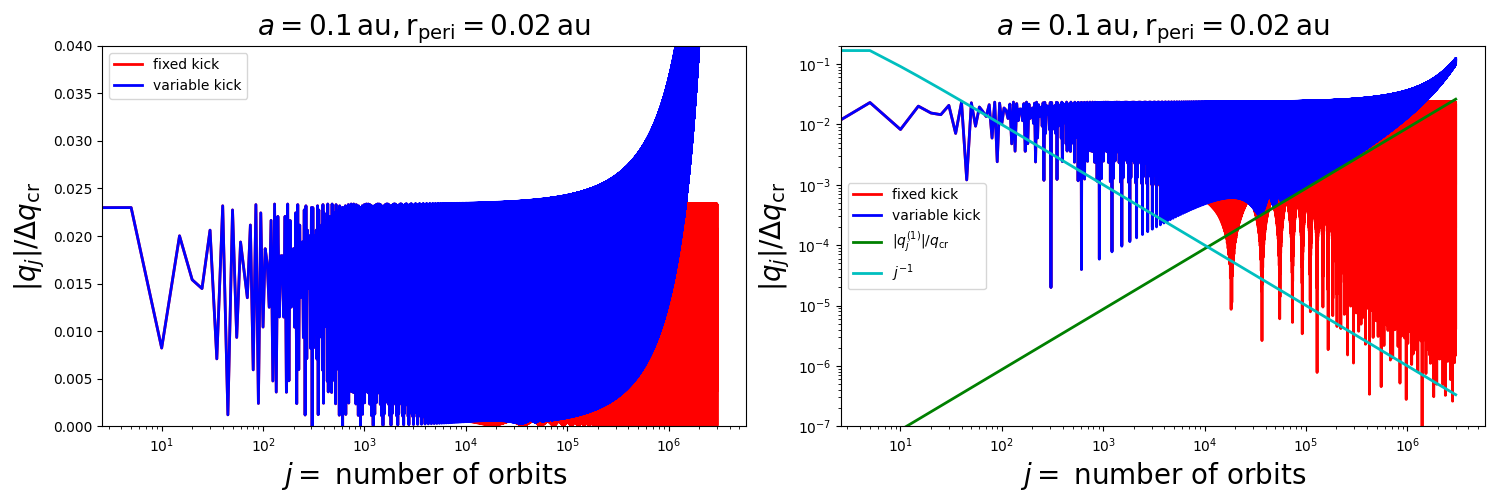} 
    \caption{Same as Figure \ref{fig:q_vs_t_1AU_0.02AU} but for $a_0=0.1\, \rm AU$. In this case, the kick size is even smaller, relative to the critical kick size, causing the growth time to increase. } 
    \label{fig:q_vs_t_0p1AU_0.02AU}
\end{figure*}

Figure \ref{fig:q_vs_t_0p1AU_0.02AU} shows $a_0=0.1\, \rm AU$ and $e=0.8$, again with pericenter $a_0(1-e_0)=0.02\, \rm AU$. The orbital energy is larger by a factor of 10 as compared to the $a_0=1\, \rm AU$ case, and hence the kick amplitude $\Delta Q_a$ is  much smaller than the critical kick $\Delta Q_{\rm a, cr}$.  Nonetheless, the growth occurs in this case as well, albeit on a longer timescale. Note that even though the initial pericenter is the same as that in Figures~\ref{fig:q_vs_t_10AU_0.02AU} and \ref{fig:q_vs_t_1AU_0.02AU}, the elliptic orbit kick is smaller than the $a_0=1\, \rm AU$ case by 60\% whereas the \citetalias{Press:77} parabolic kick would have given an identical kick size.  

Growth in the mode amplitude, even for kick amplitudes much  smaller than the critical value, is the central result of this paper. The authors are unaware of any previous investigations demonstrating this small-kick growth, even though we are solving the same equations. It seems to be a region of parameter space whose importance was not recognized as longer-timescale simulations are needed. In detail, \citet{Mardling:95} used numerical integration of the equations of motion, but only for $\sim 10^2$ orbits; \citet{Vick:18} used the mapping method with fixed kick for $\sim 10^4$ orbits; and \citet{Wu:18} used numerical integration of the equations of motion for $\sim 10^3$ orbits. 

As for the case of the more rapid diffusive growth, the small-kick growth may be of astrophysical interest if large amplitudes may be obtained which then allow rapid orbital evolution.

\section{ analytic model }
\label{sec:analytic}

To better understand the origin of the growth observed for small kick size, an analytic model is developed in this section to treat the initial growth with $|Q_a| \ll 1$. The key new element is the variation of the kick size as the orbit varies. 

\subsection{ Constant kick }

To review, first consider the Vick and Lai map (\citealt{Vick:18}, hereafter \citetalias{Vick:18}; see also \citealt{Wu:18}) 
\be
Q_{j+1} & =& \left( Q_j + \Delta Q \right) e^{-i \omega_a P_0 \left( 1 + |Q_j+\Delta Q|^2 \right)^{-3/2} }
\label{eq:vlmap}
\ee
which assumes constant kick $\Delta Q$. Here $Q_j$ is the mode amplitude and $j$ labels the orbit number for the mapping.
The initial condition is taken to be $Q_0=0$ here. In the large kick limit, the phase factor in the exponent is effectively random giving
$|Q_j| \simeq |\Delta Q|j^{1/2}$ \citep{Kochanek:92,Mardling:95}.
For the small kick limit, an analytic solution has previously been derived by \citetalias{Vick:18}. Treating the orbital period $P_{j+1} = P_0 \left( 1 + |Q_j+\Delta Q|^2 \right)^{-3/2} \simeq P_0$, for small $Q_j$,  the resulting linear difference equation may be solved using a ``summing factor" \citep{1978amms.book.....B}, giving an the oscillatory solution
\be
Q_j^{(0)} & = & \Delta Q \left( \frac{ e^{-ij \omega_a P_0} - 1}{1 - e^{i\omega_a P_0} } \right)
\label{eq:oscillatory}
\ee
with energy
\be
|Q_j^{(0)}|^2 &= & 
  |\Delta Q|^2 \left(  \frac{ \sin^2 (j\omega_a P_0/2)}{\sin^2(\omega_a P_0/2)} \right).
\ee
This is analogous to the driven, undamped harmonic oscillator, and implies purely oscillatory behavior with no growth. 
Equation \ref{eq:oscillatory} breaks down for resonances $\omega_a P_0 = 2\pi n$ for integer $n$ (not to be confused with mean motion). For an exact resonance, there is an initial growth phase linear in time, $Q_j \simeq j \Delta Q$, which terminates when the frequency shifts at $\omega_a P_0 j^2 |\Delta Q|^2 \ga 1$ (e.g. VL18). 
At this point, the amplitude is large enough to significantly shift $\omega_a P/\pi$ away from an integer value, and the linear growth transitions into oscillation.

The effect of varying orbital period may be included in the \citetalias{Vick:18} map assuming that $\omega_a P_0 |\Delta Q|^2 \ll 1$ so that the changes in orbital period is small. The exponent in Equation \ref{eq:vlmap} may then be expanded as 
\be
Q_{j+1} & \simeq & \left( Q_j + \Delta Q \right) e^{-i \omega_a P_0}
\nonumber \\ & \times & 
\left( 1 + \frac{3}{2} i \omega_a P_0 |Q_j+\Delta Q|^2 \right).
\ee
This equation may be solved using perturbation theory, $Q_j \simeq Q_j^{(0)}+Q_j^{(1)}$, with Equation \ref{eq:oscillatory} as the zeroth order solution. The correction $Q_j^{(1)}$ has a piece linear in $j$, as well as oscillatory pieces ringing at different frequencies sourced by the nonlinear term. The linear in $j$ term is
\be
Q_j^{(1)} & \simeq & \frac{9}{8} i \omega_a P_0 j\, \left( \frac{|\Delta Q|^2}{\sin^2(\omega_aP_0/2)}\right) 
\nonumber \\ & \times & 
\left(  \frac{\Delta Q}{ 1-e^{i \omega_a P_0}} \right) e^{-ij\omega_a P_0}.
\label{eq:Qj1_varying_period}
\ee
The expansion $Q_j^{(1)} \la Q_j^{(0)}$ is only valid for small $j$. However the form of Equation \ref{eq:oscillatory} and \ref{eq:Qj1_varying_period} suggest that these two terms can be combined to yield a purely oscillatory solution $Q_j^{(0)}+Q_j^{(1)} \propto \exp(-ij\omega_a P)$ valid at all times, but with a slightly shifted period
\be
P & \simeq P_0 \left( 1 - \frac{9}{8}\, \frac{ \displaystyle |\Delta Q|^2 }{\displaystyle \sin^2(\omega_a P_0/2)} \right).
\ee
Hence in the constant-kick case, and for small kick sizes $\Delta Q \la \Delta Q_{\rm cr}$, the mode amplitude is expected to have purely oscillatory behavior at all times, ringing at multiples of the orbital period $P$ that is shifted from the zero-amplitude period $P_0$. This purely oscillatory behavior agrees with the constant-kick results in Figures \ref{fig:q_vs_t_1AU_0.02AU} and \ref{fig:q_vs_t_0p1AU_0.02AU} over $\ga 10^6$ orbits, and we see the same dependence over even longer timescales in runs not shown here.

\subsection{ Dependence of kick size on amplitude}

With the constant kick case understood analytically, we now turn to the role of  kick varying from orbit to orbit. With the growth of the mode amplitude $Q$, the semi-major axis and eccentricity vary as in Equations \ref{eq:da} and \ref{eq:de}. As $a$ and $e$ vary, so does the kick $\Delta Q$. For elliptic orbits, and mode frequency $\omega_a$ larger than pericenter-crossing frequency $n(1-e)^{-3/2}$, the main dependence of $\Delta Q$ on $a$ and $e$ is through the exponent (see Equation \ref{eq:Klm_exponent} and Equation \ref{eq:hansen_asymp_approx} in the Appendix)
\be
\Delta Q & \propto & e^{-(\omega_a/n)(\cosh^{-1}(1/e) - \sqrt{1-e^2})}
\nonumber \\  
&  \equiv & e^{-(\omega_a/n) g(e)}.
\ee
The kick $\Delta Q$ at finite $Q$ can then be related to the kick at $Q=0$, $\Delta Q_0$, using Equations
\ref{eq:a_of_Q} and \ref{eq:e_of_Q} to find
\be
\Delta Q & \simeq \Delta Q_0 \left( 1 - \zeta |Q|^2 \right),
\label{eq:kick_variation}
\ee
where derivatives of the exponent give
\be
\zeta & = & \left( \frac{\omega_a}{n_0} \right) \left( - \frac{3}{2} g - \frac{dg}{de} \frac{1-e^2}{2e}
+ \frac{dg}{de} \frac{\sqrt{1-e^2}}{2e} \frac{m_a n_0}{\omega_a} \right).
\ee
In the $1-e \ll 1$ limit this reduces to
\be
\zeta & \simeq & \left( \frac{\omega_a}{n_0} \right) \left( 1 - e_0 \right)^{5/2} 
\left( \frac{4\sqrt{2}}{5} - \frac{m_a n_0}{\omega_a (1-e_0)^{3/2}} \right)
\nonumber \\ & \simeq & 
\left( 1-e_0 \right) 
\left[ \frac{16}{5} \left( \frac{\omega_a}{\omega_{\rm dyn, 1}}\right)\left( \frac{r_{\rm p}}{r_{\rm p, Roche 1}} \right)^{3/2} 
- m_a 
\right].
\label{eq:zetaapprox}
\ee
To make the numerical estimate in the second expression, let body 1 be a planet and body 2 a much more massive star. Here $\omega_{\rm dyn,1} = \sqrt{ GM_1/R_1^3}$ is the dynamical frequency of the planet, and $r_{\rm p, Roche 1} \simeq 2R_1 (M_2/M_1)^{3/2}$ is the orbital separation at which the planet fills its Roche lobe. 
For an f-mode in the planet with $\omega_a \sim \omega_{\rm dyn, 1}$, and the planet nearly filling its Roche lobe at pericenter, the two terms in Equation \ref{eq:zetaapprox} are comparable for $m_a=2$. For wider orbital separations the first term will dominate.  
For a fixed pericenter distance $r_p$, the $1-e_0$ factor out front implies $\zeta$ is smaller for more eccentric orbits.

\subsection{ Map including varying kick }

Equation \ref{eq:kick_variation} suggests the following new map that allows for a variable kick size:
\be
Q_{j+1} & =& \left[ Q_j + \Delta Q_0\left( 1 - \zeta |Q_j|^2\right) \right] 
\nonumber \\ & \times & 
e^{-i \omega_a P_0 \left( 1 + \left| Q_j + \Delta Q_0 \left( 1 - \zeta |Q_j|^2\right) \right|^2 \right)^{-3/2} }.
\label{eq:newmap}
\ee
Naively, one would expect that the inclusion of the varying kick could only make the mode amplitudes smaller, as the kick decreases with increasing mode amplitude. However, the introduction of the $\zeta |Q_j|^2$ term in the prefactor gives rise to a resonance and linear growth in the mode amplitude, as will now be shown.

Equation \ref{eq:newmap} will be solved using perturbation theory in powers of the mode amplitude.
At zeroth order, variations in  orbital period 
and kick size  are ignored to find Equation \ref{eq:oscillatory}.
As the shift in orbital period has already been discussed, here we set the exponent to $-i \omega_a P_0$ and focus on the change in kick size in the prefactor. The deviation from $Q_j^{(0)}$, again called $Q_j^{(1)}$, satisfies the equation
\be
  Q_{j+1}^{(1)} & = & 
 \left[  Q_{j}^{(1)}  -\zeta \Delta Q_0 |Q_j^{(0)}|^2 \right] e^{-i \omega_a P_0 }
 \nonumber \\ & = & 
 \left[ Q_{j}^{(1)}  -\zeta \Delta Q_0
 |\Delta Q_0|^2 \left( \frac{2 - e^{-ij \omega_a P_0} - e^{ij \omega_a P_0} }{ 4\sin^2(\omega_a P_0/2) } \right) 
 \right] 
  \nonumber \\ & \times & 
e^{-i \omega_a P_0 }.
 \ee
 Multiplying by the summing factor $\exp(i(j+1)\omega_a P_0)$ puts the equation in the form
 \be
 && Q_{j+1}^{(1)}e^{i(j+1)\omega_a P_0}
 - Q_{j}^{(1)}e^{ij\omega_a P_0}
   \nonumber \\ &= & 
  \left( \frac{\zeta \Delta Q_0 |\Delta Q_0|^2}{4 \sin^2(\omega_a P_0/2)} \right)
\left( 1 - 2 e^{ij\omega_a P_0} + e^{2ij\omega_a P_0} \right),
 \ee
 where the left hand side is now a discrete derivative which can be immediately summed \citep{1978amms.book.....B}. On the right-hand side, the constant term gives $j$ upon summing. This is the resonant term. The other two terms are oscillatory upon summing, and will be ignored as they do not grow. The linearly growing piece of the solution is then
\be
Q_{j}^{(1)}  & \simeq & \left( \frac{ \zeta \Delta Q_0 |\Delta Q_0|^2 }{ 4 \sin^2(\omega_a P_0/2)}\right)
\, j \, e^{- ij\omega_a P_0}.
\label{eq:Qk1}
\ee
The explicit factor of $j$ implies the amplitude increases linearly in time $t=jP_0$, while the oscillatory piece is the same as for $Q_j^{(0)}$. Here $Q_j^{(1)}$ cannot be combined with $Q_j^{(0)}$ to give a period shift as it does not have the right complex phase. Equation \ref{eq:Qk1} demonstrates growth at small amplitudes in the analytic model.

\begin{figure}[htb]
    \epsscale{1.1}
    \plotone{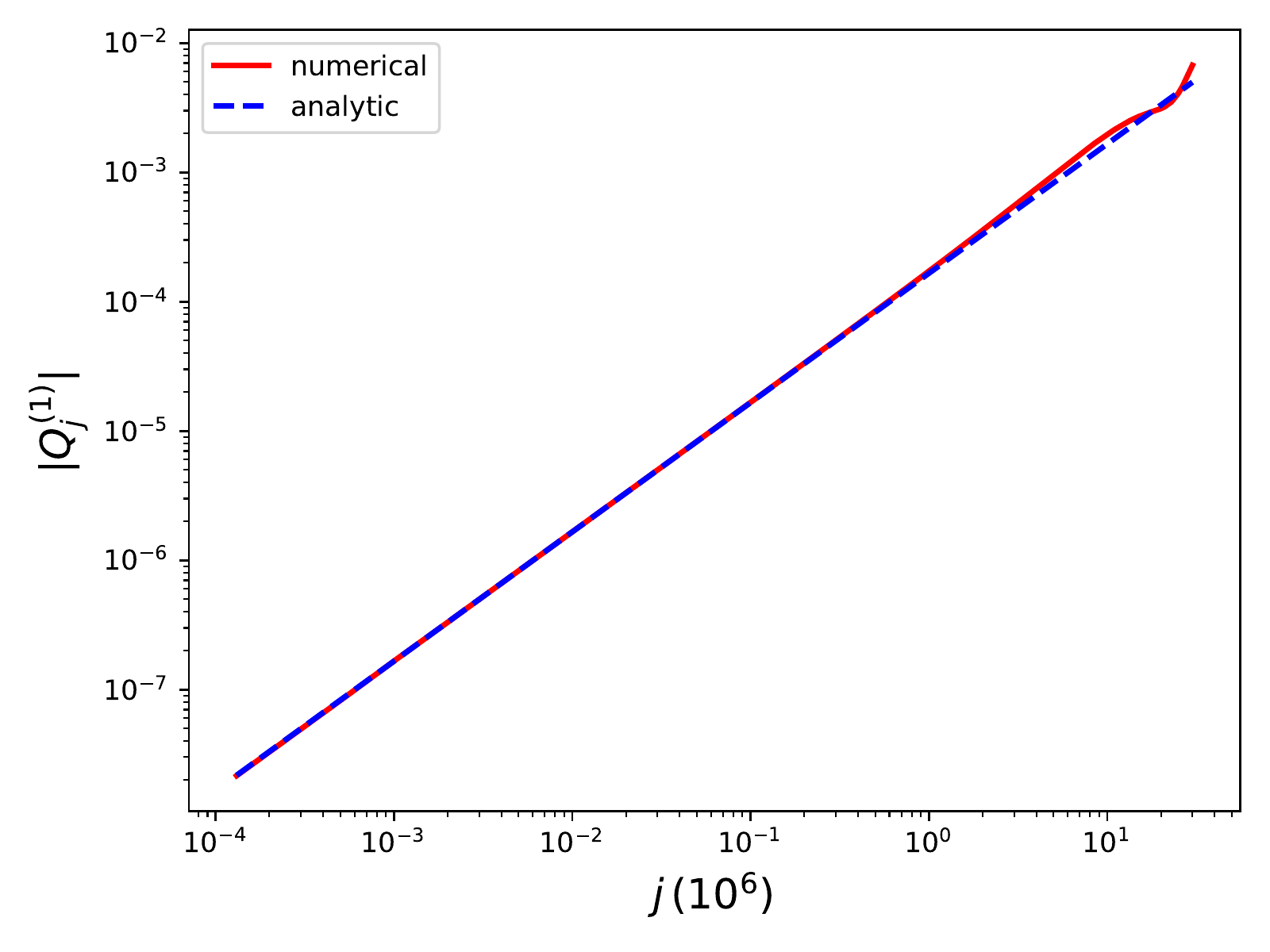}
    \caption{Mode amplitude $|Q^{(1)}_j|$ versus orbit number $j$ for the analytic model in Equation \ref{eq:newmap}. The variation of the orbital period in the exponent has been ignored. The red line is the numerical solution minus $Q^{(0)}_j$ from Equation \ref{eq:oscillatory}. The blue line is the linearly increasing term in Equation \ref{eq:Qk1}. The parameters used are $\Delta Q_0=4\times 10^{-4}=10^{-2}\Delta Q_{\rm cr}$, $\omega_a P_0 = 2\pi \times 100.1$, and $\zeta=1$. Only roughly 1 in 100 points are plotted.} 
    \label{fig:Q1_vs_k_analytic}
\end{figure}

Figure \ref{fig:Q1_vs_k_analytic} shows the numerical solution of Equation \ref{eq:newmap}, with $Q^{(0)}_j$ from Equation \ref{eq:oscillatory} subtracted off, versus the linearly increasing solution in Equation \ref{eq:Qk1}. The orbital period has been treated as constant so that the exponent is $-i\omega_a P_0$. There is good agreement with the analytic solution over millions of orbits and amplitude growth by a factor of $\sim 10^6$, verifying that resonant growth due to the varying kick is the cause of the amplitude increase for small kick size.

The linear in time growth is seen not only for this simplified analytic model, but also for the ``exact" map in Equations \ref{eq:step1}-\ref{eq:step3}. The right-hand panels in Figures \ref{fig:q_vs_t_1AU_0.02AU} and \ref{fig:q_vs_t_0p1AU_0.02AU} show the numerical solutions against the analytic solution from Equation \ref{eq:Qk1}, where the parameters in that equation such as $P_0$ and $\zeta$ have been computed using the initial values of $a$ and $e$. While the lower envelope of the constant kick line decreases with time, as the $\sin$ function $\sin(j\omega_a P)$ is effectively sampled with random arguments that are occasionally near $0 $ and $\pi$, the case with varying kick shows a lower envelope increasing linearly in time, and with good agreement with Equation \ref{eq:Qk1}. Hence while Equation \ref{eq:Qk1} was derived for a toy model, it appears to have qualitative and some quantitative agreement with the exact map.

\subsection{ Growth time in the small kick regime }


The orbital evolution is fastest when the mode amplitude is largest. A fiducial growth time to reach the wave-breaking amplitude $|q_a|=q_{\rm max}$ is given by
\be
\frac{t_{\rm gr}}{P_0} &= &  2^{1/2} q_{\rm max}\left( \frac{M_1}{M_2} \right)^{1/2} \left( \frac{a_0}{R_1} \right)^{1/2} \left( \omega_a P_0 \right)^{3/2} 
   \nonumber \\ & \times & 
\left( \frac{\Delta Q_{\rm cr}}{|\Delta Q_0|}\right)^3  \left( \frac{ 4 \sin^2(\omega_a P_0/2)}{ \zeta }\right)
\nonumber \\ & \simeq &  
10^5\ \left( \frac{q_{\rm max}}{0.1} \right)
\left( \frac{M_1}{10^{-3}\, M_2} \right)^{1/2} \left( \frac{a_0}{200\, R_1} \right)^{1/2}
   \nonumber \\ & \times & 
\left( \frac{\omega_a} { \sqrt{GM_J/R_J^3} }\right)^{3/2}
\left( \frac{P_0}{1\, \rm yr} \right)^{3/2}
   \nonumber \\ & \times & 
\left( \frac{\Delta Q_{\rm cr}}{|\Delta Q_0|}\right)^3  \left( \frac{ \sin^2(\omega_a P_0/2)}{ 10\zeta }\right).
\label{eq:tgr}
\ee

Midway between resonances, $\omega_a P_0 \gg 1$ and $\zeta \la 1$ imply growth times much longer than the orbital time. The minimum growth time is found by using the maximum kick size $Q_{\rm cr}$ for the small-kick regime, and depends on the orbital separation and mass ratio.   For the fiducial parameters used, and kick sizes approaching the critical value, growth times $\sim 10^6$ orbits are much longer than the orbital timescale, but may also be much shorter than the ages for systems that have undergone billions of orbits.

In this idealized dynamical system, {\it all} binaries whose response to the tidal force may be modeled as an undamped oscillation will undergo growth, either in the small or large kick regime. In practice, the growth time may be so long as to be uninteresting when other effects are taken into account. 
Growth may also be limited by mode damping, as discussed in the following section.

\section{ the role of linear damping }
\label{sec:damping}

Including linear damping in the map, but again ignoring variations of the orbital period in the exponent, Equation \ref{eq:newmap} becomes
\be
Q_{j+1} & =& 
 \left[ Q_j + \Delta Q\left( 1 - \zeta |Q_j|^2\right) \right] e^{-(i \omega_a +\gamma) P_0}.
\label{eq:damping}
\ee
At zeroth order in the amplitude the solution becomes
\be
Q_j^{(0)} & = & \Delta Q_0 \left( 
\frac{ e^{-j\gamma P_0 - ij\omega_a P_0} - 1 }
{1 - e^{\gamma P_0 + i \omega_a P_0} }
\right)
\ee
so that the oscillations damp away for $j \gamma P_0 \ga 1$ and the zeroth order piece becomes independent of time. At first order there is also a constant term, as well as the linearly increasing term which becomes
\be
Q_{j}^{(1)}  & \simeq & \left( \frac{ \zeta \Delta Q_0 |\Delta Q_0|^2 }{ 4 \sin^2(\omega_a P_0/2) + (\gamma P_0)^2}\right)
   \nonumber \\ & \times & 
\, j \, e^{-j\gamma P_0 - ij\omega_a P_0}.
\label{eq:Qk1damp}
\ee
The denominator takes on the expected Lorentzian form when damping is included, and an exponential damping in time is also found. The time-dependence $Q_j^{(1)} \propto j \exp(-j\gamma P_0)$ implies a maximum amplitude near the time $jP_0 \sim \gamma^{-1}$. In order for the mode amplitude to have time to grow to large amplitudes  requires  that
\be
t_{\rm gr} & \la & \gamma^{-1}.
\ee
Growth then requires long mode damping times and large kick sizes, which in the small kick regime are limited by the critical value.

\begin{figure}[htb]
    \epsscale{1.1}
    \plotone{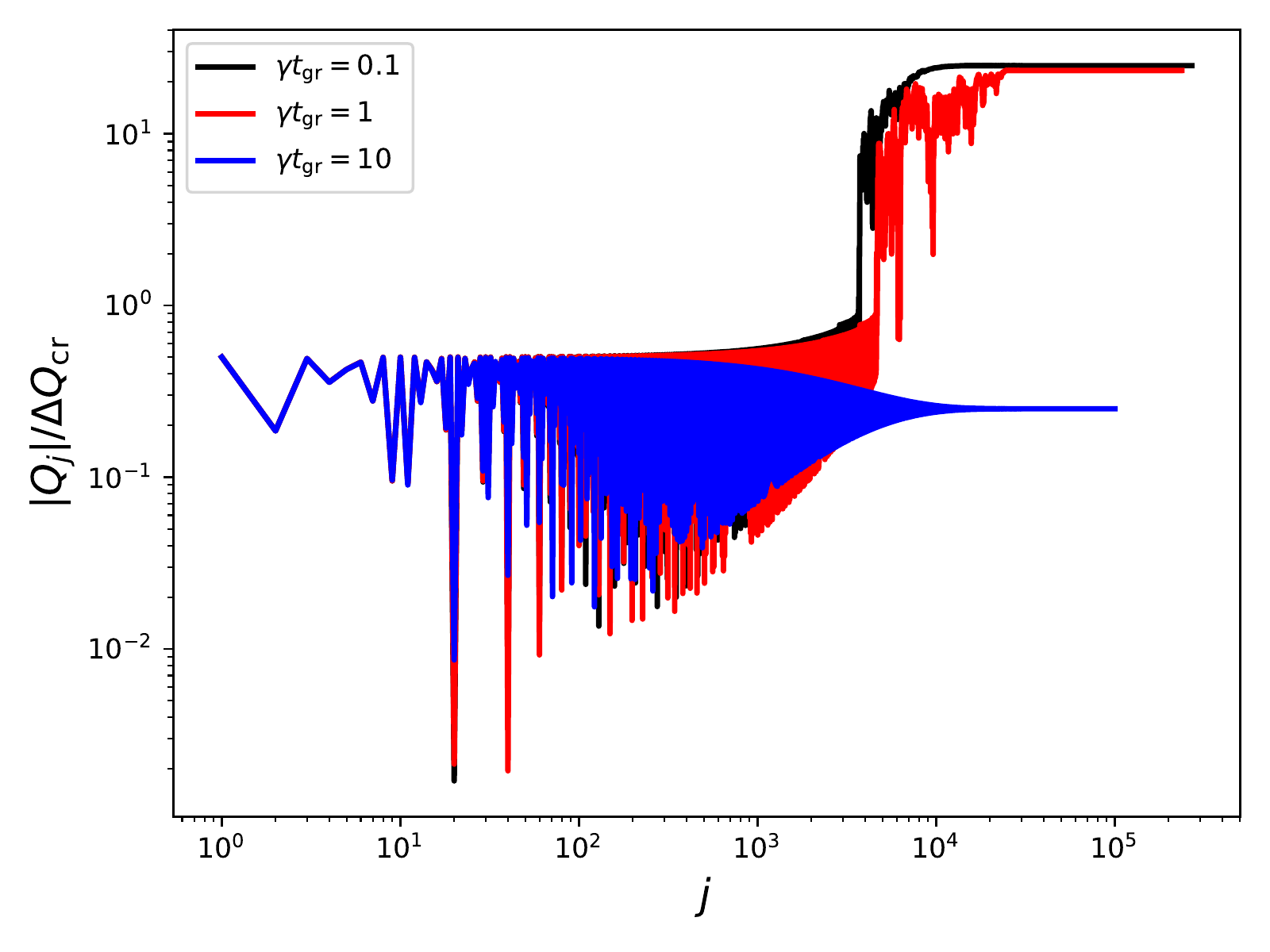} 
    \caption{Mode amplitude versus orbit number for the map in Equation \ref{eq:damping} including varying kick size and damping. The lines represent solutions to Equation \ref{eq:newmap} including both damping and orbital period changes in the exponent. The parameters used are $\omega_a P_0 = 2\pi \times 100.5$, $\zeta=1$ and initial kick $\Delta Q_0=0.5\Delta Q_{\rm cr}$. The three lines represent different values of damping given by $\gamma t_{\rm gr}=0.1,1,10$, where $t_{\rm gr}$ is in Equation \ref{eq:tgr}. } 
    \label{fig:damping}
\end{figure}

The analytic expectations for the role of damping are verified in Figure \ref{fig:damping}, in which the new map is used including mode damping as well as orbital period changes in the exponent. Damping rates $\gamma t_{\rm gr} \gg 1$ (blue line) are unable to grow sufficiently to shift the frequency and start the phase of more rapid diffusive growth. In this case the amplitude settles down to a constant. For the weakly damped case $\gamma t_{\rm gr} =0.1$ (black line), the amplitude grows linearly until diffusive growth commences. The saturation in this model is artificial, and is enforced by the kick going to zero when $|Q_j|^2$ approaches $\zeta^{-1}$. The case of $\gamma t_{\rm gr}=1$ (red line) also shows growth, although it is delayed as compared to the weakly damping case.

With the origin of the linear growth understood, the next section will discuss an application to hot Jupiters.

\section{ numerical results for hot-Jupiters}
\label{sec:results}

The central result of this paper is that oscillation modes in inviscid bodies in an eccentric binary will generically grow.  The question addressed in this section is the extent to which mode growth in the slower small kick regime causes further orbital evolution outside the diffusive regime.

The growth must saturate at some point, and the orbital evolution rate is sensitive to the saturation amplitude. \citet{Wu:18} suggested that mode amplitudes would be limited below some $q_{\rm max}$ by catastrophic nonlinear processes, and would be reduced during a sudden transition to a much smaller amplitude $q_{\rm min}$, after which normal evolution by the mapping method would resume. Cycles of mode growth and sudden decrease then occur. For the diffusive case, this orbital evolution ends when the mode has 
$|\Delta Q_0| \la \Delta Q_{ cr}$ 
as well as $|Q|<Q_{\rm cr}$ and growth no longer occurs. 


Here we consider a Jupiter mass and radius planet orbiting a solar mass perturber. The initial condition for the orbit is $a=3\, \rm AU$ and  $e=0.995$. Only the prograde f-mode in the planet is evolved, and it is assumed to have frequency $\omega_a=5.64\times 10^{-4}\, \rm rad\, s^{-1}$ and overlap integral $I_{a\ell m}=0.4$ \citep{Yu:21}. The effect of tidal heating on the planetary structure is not included. The ``exact" mapping method discussed in Section \ref{sec:statement} is used for the map.  The angular momentum of the mode is taken into account, so that both orbital energy and angular momentum vary.  As discussed in the Appendix, the dynamical tide piece of the one-kick integrals $K_{\ell m}(k,e)$ are interpolated over a fine grid in $k$ and $e$, and the equilibrium tide contribution is ignored.

\begin{figure*}[htb]
    \epsscale{1.0}
    \plotone{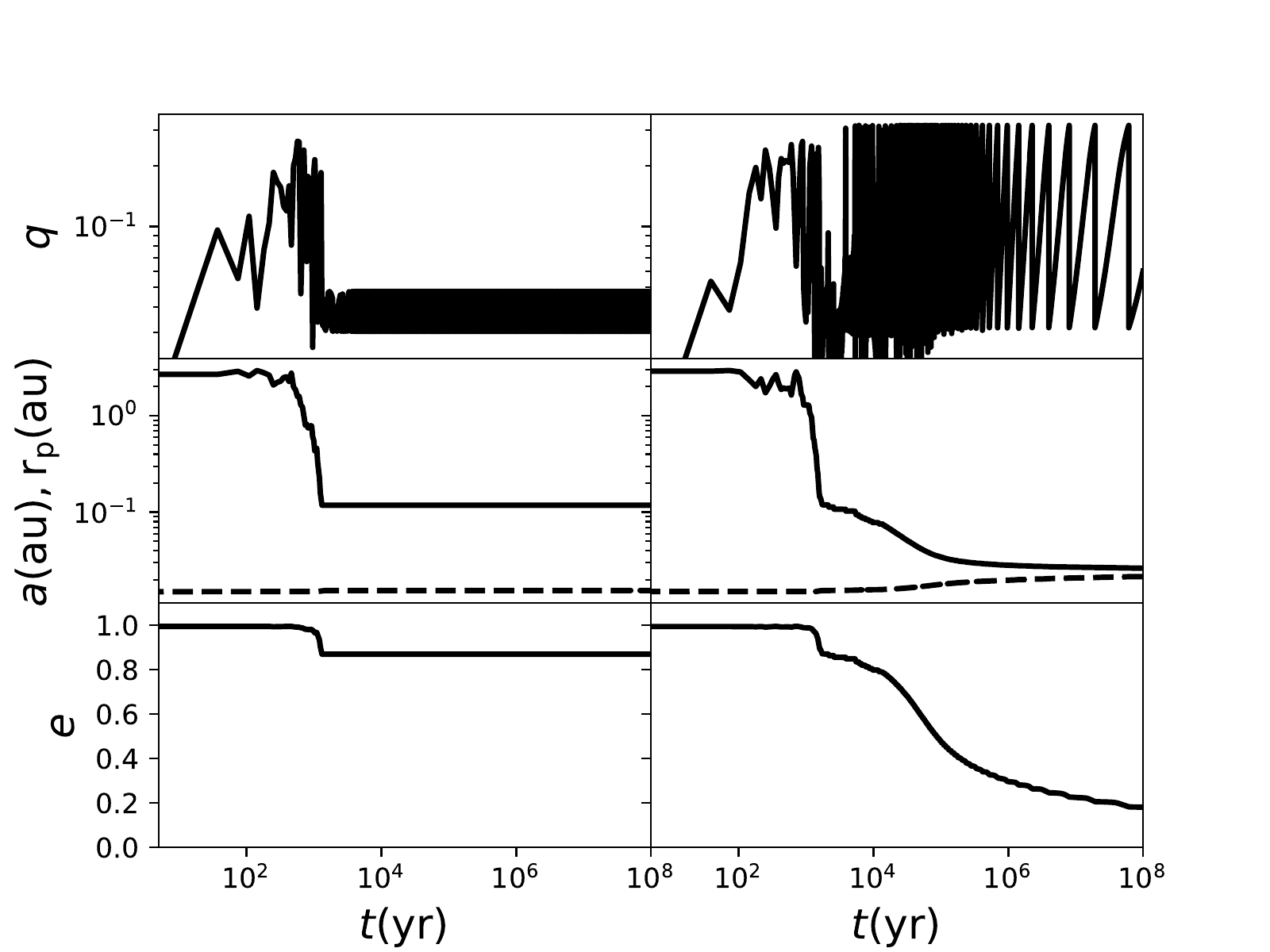} 
    \caption{Evolutionary calculation for a binary containing a hot Jupiter perturbed by a solar-type star. In the left column the kick amplitude is fixed throughout the evolution at the initial value. The kick amplitude is allowed to vary self-consistently in the right column. The top, middle, and bottom panels show mode amplitude, semi-major axis and pericenter distance, and eccentricity versus time. The values $q_{\rm min}^2=10^{-3}$ and $q_{\rm max}^2=0.1$ have been used, as well as a single prograde ($m=2$) mode. Note that the orbital evolution halts after $\sim 10^4\, \rm yr$ for fixed kick amplitude. Due to the large number of mapping iterations, points are only plotted over a logarithmically increasing time step, so that some detail may be missing over short time intervals.  }  
    \label{fig:hj12}
\end{figure*}

The left column of Figure \ref{fig:hj12} shows evolution if the kick amplitude is fixed during the evolution and if there is no linear damping. For the assumed parameters, the pericenter distance is small enough that the evolution starts out in the chaotic regime and growth to $q_{\rm max}$ occurs in $\simeq 10^3\, \rm yr$. The resulting large amplitude causes a significant decrease of the semi-major axis to $a \simeq 0.1\, \rm AU$ and eccentricity $e \simeq 0.9$. At that point, the kicks are no longer large enough for the evolution to stay in the chaotic regime, due to the increase in orbital binding energy as compared to mode energy. Once the system drops out of the diffusive regime, the orbital evolution halts.

The right column of Figure \ref{fig:hj12} shows a simulation for the same parameters as the left column, but allowing the kick size to vary self-consistently as the orbit varies, in particular as the pericenter separation varies. The key difference is that even after the system is no longer in the diffusive regime ($t \ga 10^3\, \rm yr$, $a \la 0.1\, \rm AU$), orbital evolution continues due to the small-kick growth. Cycles of small-kick growth and rapid decay continue to occur. As the pericenter  moves outward, the kick size decreases with time and this causes the small kick growth timescale to increase with time, so that the duration of each ``sawtooth" cycle gets longer. Within $t\simeq 10^7\, \rm yr$ the semi-major axis further shrunk to $a \simeq 0.03\, \rm AU$ and, significantly, the eccentricity to $e \simeq 0.2$. While this eccentricity is still larger than most present-day hot Jupiters, the orbit is now so close and nearly circular that the dissipative equilibrium tide may be effective in further circularizing the orbit (e.g. \citealt{Ivanov:04}). Hence the combination of an early phase of diffusive evolution, and a subsequent phase of small kick evolution may achieve better agreement with the observed population of hot Jupiters.

\begin{figure*}[htb]
    \epsscale{1.0}
    \plotone{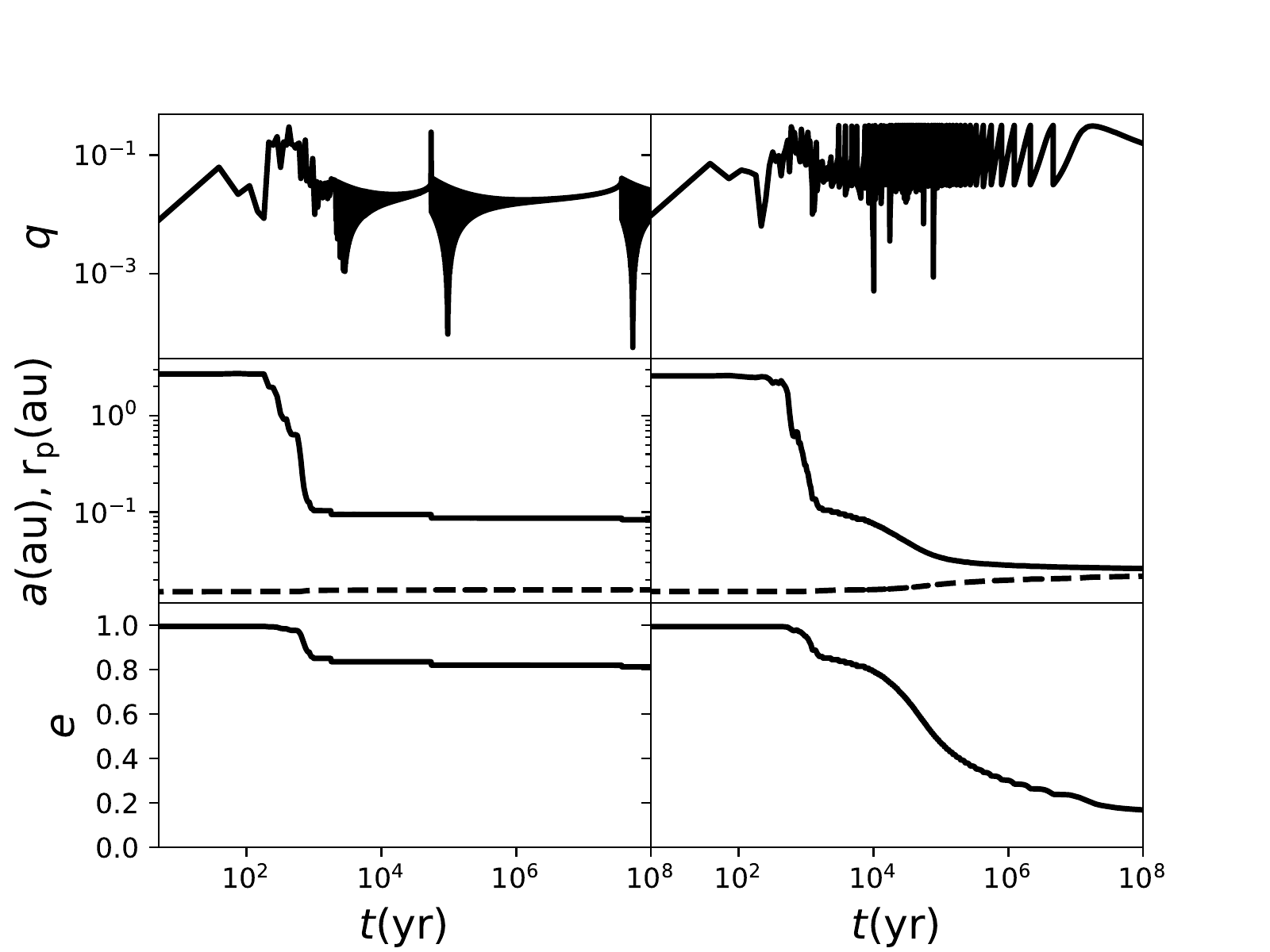} 
    \caption{Same as Figure \ref{fig:hj12}, but including an f-mode damping rate $\gamma(t) = 1/t$ that decreases as the planet ages.  }  
    \label{fig:hj_56}
\end{figure*}

Figure \ref{fig:hj_56} considers the same parameters and initial conditions as Figure \ref{fig:hj12}, but now includes a specific model for mode damping in which the damping rate $\gamma(t)=1/t$ is used, where $t$ is the age. This damping rate rapidly decreases over time. Again there is an initial phase of diffusive growth seen in both panels, and the evolution in the small-kick phase is both qualitatively and quantitatively similar. In particular, the orbit again  circularizes to $e \sim 0.2$ within $10^7\, \rm yr$. Some details of the mode amplitude evolution are different, in particular specific times where the amplitudes deviate from the $\gamma_a=0$ curves, likely due to passing through resonances and then undergoing subsequent damping.
In addition, at late times $t=10^7-10^8\, \rm yr$, the mode growth is slower, and starts decaying near the end. This may be due to the kick size becoming small enough that the growth time is comparable to the mode damping time.

Is the assumption of negligible mode damping appropriate for the Jupiter-size planets of interest? If we assume that mode damping times are comparable to the age of the planet, perhaps after an initial thermal time \citep{2006ApJ...650..394A}, then at some point the age will be greater than the small-kick growth time, and growth will begin to occur. For example, if their linear damping is dominated by turbulent viscosity in the interior convection zone, the damping rate is proportional to a power of the internal luminosity through the eddy speed. For instance, for eddy viscosity reduced by a factor $(\omega_a \tau_{\rm eddy})^{-2}$, where $\tau_{\rm eddy}$ is the eddy turnover time, turbulent viscosity is proportional to luminosity \citep{Goldreich:77}. After formation, as the planet cools and the internal luminosity drops, the damping rates decrease as well. \citet{2006ApJ...650..394A} shows scaling for power-law indices of luminosity with time which are order unity, which motivates the scaling used here. We have constructed MESA models for $M=0.5-2\, M_{\rm J}$ planets, and for quadratic viscosity (appropriate for short mode periods) we do find damping rates falling steeply with age, motivating the simple model adopted here.

\section{Summary and Conclusions}
\label{sec:summary}

The goal of this paper is to understand the long-term evolution of a binary orbit including the dynamical tide. As integrations of the coupled orbital and mode equations over billions of orbits are challenging, the mapping method is used \citep{Ivanov:04,Vick:18, Wu:18}, in which the dynamical tide ``kick" at pericenter causes an exchange of energy and angular momentum with the orbit, which may be modeled analytically instead of through direction integration of the equations of motion.

We have improved upon previous investigations by using the dynamical tide kick for a bound Keplerian orbit (see the Appendix). Long term simulations in which the kick was allowed to vary self-consistently (Section \ref{sec:growth}) gave the surprising result that even for small kick amplitudes the mode amplitude grew to large values. Previous investigations \citep{Mardling:95, Vick:18, Wu:18} using a fixed kick or shorter integration times  did not show this growth for small kick values. Hence, the central result of this paper is that, for weak mode damping, the mode amplitude will grow to large values for any values of semi-major axis, eccentricity, and kick size, as long as the kick is allowed to vary self-consistently as the orbit varies. 

When mode damping is included it is found that this can suppress the small-kick regime of growth if the small-kick growth time is longer than the mode damping time. As f-mode damping times in Jupiter-size planets increase strongly with age, for instance if the damping is due to turbulent viscosity, then for sufficiently close orbits eventually an age will be reached where the small-kick growth may occur.

The growth of the mode amplitude must eventually saturate, and the saturation amplitude effects the orbital evolution as it is most rapid when the mode amplitude is largest. We used the nonlinear saturation prescription from \citet{Wu:18} in which the amplitude is decreased when it becomes larger than a critical value. The resultant small-kick evolution was then a ``sawtooth" pattern of linear growth and sudden decreases. As the orbit evolves and the pericenter moves out, the kick size decreases and the interval for the sawtooth becomes longer. If this continued until the growth time was longer than the damping time, the evolution would shut off, but could perhaps restart once these two timescales again become comparable.


We have performed preliminary estimates of the dynamical tide kicks to the mode amplitude for the observed transiting exoplanets. We find a  sample of systems with estimated small-kick growth times (see Equation \ref{eq:tgr}) much shorter than the Gyr ages of the systems. While this is suggestive that the mode amplitudes in these systems may be undergoing growth, if they are not damped too heavily, the circularization time is much longer than the growth time, and depends on the prescription for saturation. We leave more detailed studies of orbital evolution of individual systems to a future investigation.

Previous investigations (e.g. \citealt{Wu:18}) found that diffusive growth of the planetary f-mode in the large-kick regime would lead to rapid orbital shrinkage, but upon exiting the diffusive regime at $e \sim 0.9$ the theory would predict a large population of highly eccentric orbits. Simulations presented here show that subsequent orbital evolution relying on the small-kick regime may further decrease the eccentricity to $e \sim 0.2$ on timescales much less than the Gyrs ages of these systems. As the equilibrium tide circularization timescales for small eccentricities may be less than the Gyr ages, that separate mechanism may take over at $e\sim 0.2$ and explain the present-day $e\sim 0$ 
 population.

There are key issues related to the evolution of the background planet which have been ignored in this study for simplicity. As the orbital evolution rate is faster for larger mode amplitude, it depends sensitively on the saturation level of the mode. As pointed out by \citet{Wu:18}, dissipation of a mode energy of order the binding energy into heat is not a small perturbation on the planet's structure. However, as f-modes are primarily surface waves, if the dissipation occurs in a thin layer near the surface, this may lead to expansion and loss of that layer, without affecting the deeper layers of the planet much. 

As a nonlinear $m_a=2$ f-mode also carries enough angular momentum to spin the planet up to the breakup speed, the spin evolution is also important. If all the angular momentum of the prograde mode is transferred to the background planet, this could result in large shifts in the f-mode frequencies. A higher prograde f-mode frequency would give an exponentially smaller $K_{\ell m}(k,e)$ kick size as $k$ is larger \citep{Yu:22}. This would make the small-kick growth time longer. However, loss of angular momentum associated with mass loss may be able to shed some fraction of the angular momentum. The role of nonlinear evolution and outflows deserves further study.


\acknowledgments

We thank Vincent Lau and Kent Yagi for useful discussions. This work was supported by NSF grant No. AST-2054353. 
HY's work at KITP is supported by NSF PHY-1748958 and by the Simons Foundation (216179, LB).
This research has made use of the NASA Exoplanet Archive, which is operated by the California Institute of Technology, under contract with the National Aeronautics and Space Administration under the Exoplanet Exploration Program.

\software{We have used the MESA stellar evolution code \citep{Paxton:11,Paxton:13,Paxton:15,Paxton:18,Paxton:19}.}

\appendix

\label{sec:integral}

Key to the mapping method is the ability to rapidly evaluate the 1-kick amplitude $\Delta q \propto K_{\ell m}(\omega_a/n,e)$. In this Appendix, the integrals for $\ell=2$ and $m=-2,0,2$ will be evaluated using the method of steepest descent to convert the oscillatory integrand into one that decreases exponentially away from the saddle point (PT and Lai 1997). The resulting expressions will require far fewer points to attain the same accuracy, and will also allow analytic approximations to be developed. In addition, the contour integration allows a separation of the kick amplitude into equilibrium and dynamical tide parts. Integration by parts greatly simplifies the dynamical tide piece.

\begin{figure}[h] 
   \centering
   \includegraphics[width=4in]{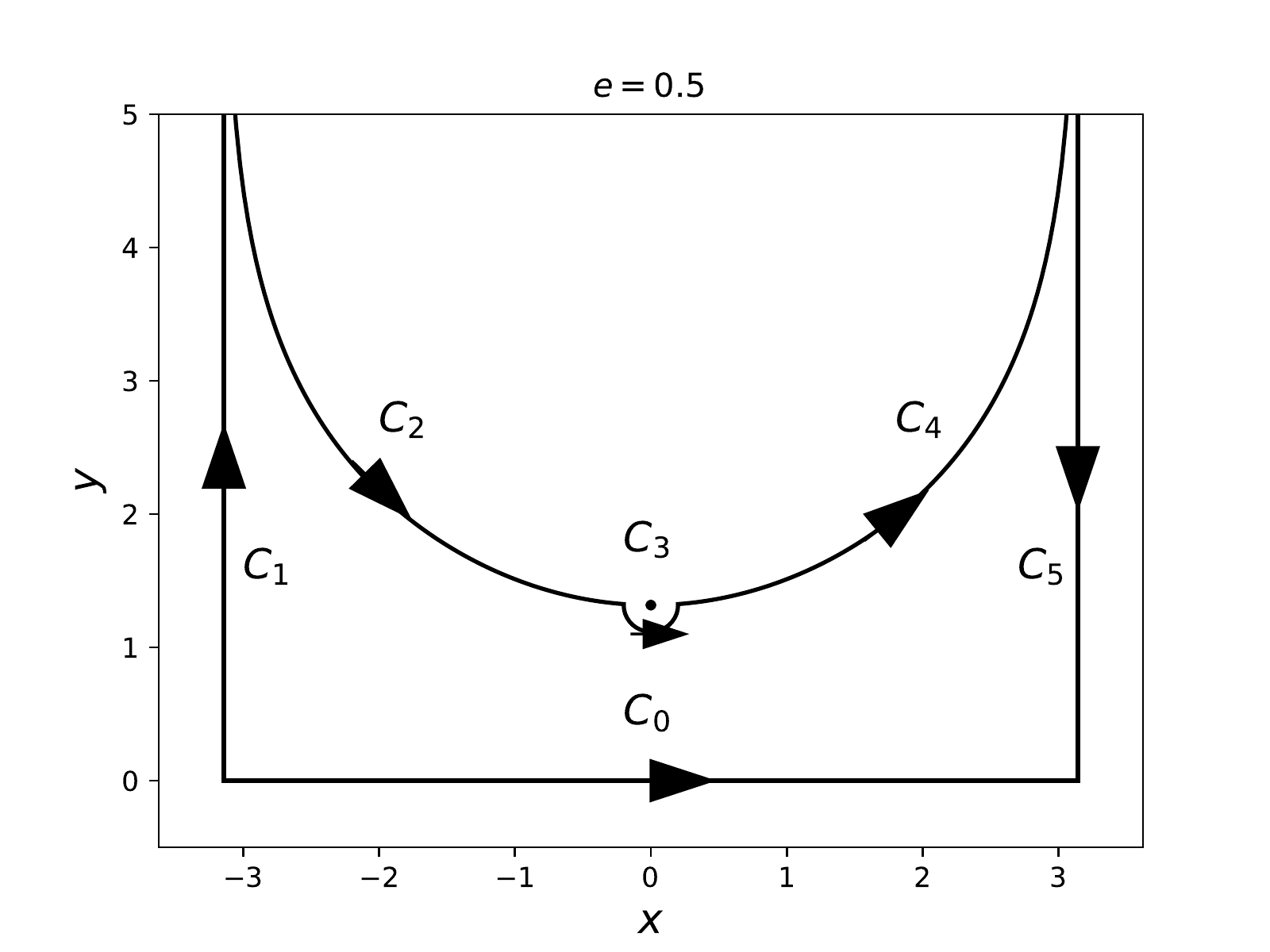} 
   \caption{Contour for the steepest descent method for the Hansen coefficients. The original integral is the contour $C_0$, which is deformed into $C_1-C_5$. The steepest descent contour is $C_2 + C_4$. The equilibrium tide contribution from the apocenter is $C_1 + C_5$, and the contribution from the semicircle $C_3$ is ero as the radius of this circle shrinks to zero.}
   \label{fig:hansen_contour}
\end{figure}

The Hansen coefficients arise in the discrete Fourier expansion in time of the tidal potential for a fixed Keplerian orbit. They are defined as
\be
X^{lm}_k(e) & = & 
\frac{1}{2\pi} \int_{-\pi}^\pi du\, \frac{ \cos \left( k(u-e\sin u) - m f(u) \right)}{(1-e \cos u)^\ell},
\label{eq:hansen_original}
\ee
where radius $r(u)/a=1-e \cos u$, true anomaly $f(u) = 2 \tan^{-1} \left( \sqrt{(1+e)/(1-e)} \tan(u/2) \right)$ and mean anomaly $nt(u) = u - e \sin u$ have been expressed in terms of the eccentric anomaly $u$. Here $\ell$, $m$, and $k$ are considered to be integers, and $k$ describes the harmonic of the orbital frequency.
The $mf(u)$ factor can be simplified using
\be
 e^{-if(u)} &= &  \frac{\cos u - e - i \sqrt{1-e^2} \sin u}{1-e \cos u}
\ee
to transform the integral to
\be
X^{\ell m}_k & = & 
 \frac{1}{2\pi} \int_{-\pi}^\pi du\  e^{i k(u-e\sin u) } 
\frac{  \left( \cos u - e - i \sqrt{1-e^2} \sin u \right)^m }
{ \left( 1-e \cos u \right)^{\ell+m} }.
\label{eq:hansen_v2}
\ee
In turning this integral over real $u$ into an integral in the complex plane, care must be taken with poles in the integrand. For $\ell + m \geq 1$, the factor $(1-e \cos u)^{-l-m}$ has infinitely many poles of order $\ell+m$ in the upper-half plane at $u=2\pi n + i y_0$, while $\cos u - e - i \sqrt{1-e^2} \sin u$ has infinitely many poles in the lower half plane at $u=2\pi n - i y_0$, where $n$ is an integer and $y_0(e) = \cosh^{-1}(1/e)$. For $k>0$, the integrand must  be closed in the upper half plane for convergence. The pole in the upper-half plane along the imaginary axis is shown in Figure \ref{fig:hansen_contour}.

The integral $K_{\ell m}(k,e)$ in Equation \ref{eq:Klm} for the one-kick amplitude is related to the Hansen coefficients as
\be
K_{\ell m}(k,e)  & = &  2\pi k (1-e)^{\ell+1} X^{\ell m}_k(e).
\label{eq:KlmXlm}
\ee
While these two integrals may look identical, there is one key difference. In the one-kick integral the ratio $k=\omega_a/n$ may be any real number, while for the Hansen coefficients $k$ is an integer. Allowing $k$ to be non-integer introduces a qualitatively different contribution to the integral, an oscillatory dependence on $k$ due to the equilibrium tide at apocenter. This contribution can actually dominate the dynamical tide piece for small eccentricity. Aside from that, the dynamical pieces will be shown to have an identical form for real or integer $k$.

The difficulty with evaluating the Hansen coefficients  for large $k$ is that small steps in $u$ are required to resolve the short-timescale oscillations in the integrand. Using complex variables,  the integration contour can be deformed to  eliminate the oscillations using the steepest descent method, producing a Gaussian integral dominated by the saddle point.

In the method of steepest descents (e.g. \citealt{1978amms.book.....B}), $k$ is considered to be large. Its  coefficient in the exponent should have constant imaginary part to eliminate oscillations. In our case, plugging in $u=x+iy$ gives
\be
ik \left( u - e \sin u \right) & = & ik \left[ x + i y - e \left( \sin x \cosh y + i \cos x \sinh y \right) \right]
\nonumber \\ & = & 
ik \left[ x - e \sin x \cosh y \right] - k \left[ y + \cos x \sinh y \right].
\ee
The imaginary part is made to be zero if we choose a steepest descent contour 
\be
\cosh y(x) & =& \frac{x}{e \sin x}, \nonumber \\
\sinh y(x) & = & + \sqrt{ \left( \frac{x}{e \sin x} \right)^2 - 1 }, 
\label{eq:contour}
\ee
where $-\pi \leq x \leq \pi$. This contour is shown in Figure \ref{fig:hansen_contour} for $C_2$ and $C_4$. The integral will be convergent in the upper-half plane for $k>0$. 
The steepest descent integrals $C_2$ and $C_4$ will be dominated by the saddle point $(x,y)=(0,y_0(e))$ as the exponential factor decreases rapidly at large $x$ 
for $k>0$ in the $y>0$ upper half plane.
In Figure \ref{fig:hansen_contour}, a closed, counterclockwise contour would not enclose any poles and so the residue is zero. Hence the contour can be deformed so that
\be
X^{lm}_k(e) & = & C_0 =  \left( C_1 + C_5 \right) +   \left( C_2  + C_3 + C_4 \right) \equiv E^{\ell m}_k(e) + D^{\ell m}_k(e)
\ee
where we have collected together two distinct terms with different behavior.

First consider the $C_1$ and $C_5=C_1^*$ contours, which will be denoted $E^{\ell m}_k(e)$. The path for $C_1$ is $u=-\pi + iy$ so that $\sin u = -i \sinh y$ and $\cos u = -\cosh y$, giving
\be
C_1 & = & \frac{i}{2\pi} e^{-ik\pi} \int_0^\infty
e^{-k(y+ke \sinh y) } 
\frac{  \left( -\cosh y - e -  \sqrt{1-e^2} \sinh y \right)^m }
{ \left( 1+e \cosh y \right)^{\ell+m} }.
\ee
For large $k$, this integral is dominated by small $y$, for which the leading order term is
\be
C_1 & \simeq & (-1)^m \frac{i}{2\pi k (1+e)^{\ell+1}} e^{-ik\pi }.
\ee
Combining with $C_5=C_1^*$ gives
\be
E^{\ell m}_k(e) & = & C_1 + C_5  \simeq (-1)^m \frac{\sin (\pi k)}{\pi k (1+e)^{\ell+1}}.
\label{eq:Elmk}
\ee
This result can be checked using the mode amplitude mapping equation. Evaluating the difference in mode amplitude from one apocenter to the next, and plugging in the equilibrium tide amplitude $q_a=U_a$ gives the equilibrium tide contribution to be
\be
K_{\ell m}(k,e) & \simeq & 2 (-1)^m \left( \frac{1-e}{1+e} \right)^{\ell+1} \sin (\pi k),
\ee 
in agreement with Equation \ref{eq:Elmk} and \ref{eq:KlmXlm}. This verifies that the $C_1+C_5$ contribution is due to the equilibrium tide contribution at apocenter. This piece is zero for integer $k$, but nonzero for general $k=\omega_a/n$. For small $e$ or large $k$ it may dominate the dynamical tide piece.

The remaining three pieces $C_2$, $C_3$ and $C_4=C_2^*$ may be greatly simplified through integration by parts. The reason is that the saddle point $u=iy_0(e)$ is also a pole, and so each of the three integrals contains divergent contributions from near the pole. These divergent pieces may be shown to all cancel each other to give the finite, physical answer. Integration by parts will remove these divergences, leaving an integrand which is finite everywhere. The boundary terms either vanish or cancel in pairs.

Integration by parts is also motivated by the fact that $X^{22}_0(e)=0$, even though the integrand is nonzero and is even about pericenter. Contributions from different portions of the integral cancel against each other. This motivates that the integrand for that case can be written as the derivative of another function, and the fundamental theorem of calculus then evaluates that other function at the endpoints, yielding zero for $X^{22}_0(e)$.

The dynamical tide contribution is labeled $D^{\ell m}_K(e) = C_2 + C_2^*$.
For $\ell=m=2$, two integrations by parts give the result
\be
2\pi C_2 & = & 
\frac{k^2}{3e^2} \int du\, e^{ik(u-e\sin u)} \left[ if(u)(1-e \cos u)  - 2ie\eta \sin u
- e \cos u ( 1 - e \cos u) 
\right. \nonumber \\ & + & \left. (1-e \cos u)\ln (1-e \cos u) + 2 \eta^2 \right],
\label{eq:hansen_22_parts}
\ee
where the integral is over $-\pi \leq x \leq 0$ along the contour $y(x)$ in Equation \ref{eq:contour}. 
Similarly, for $\ell=2$ and $m=0$, one integration by parts gives
\be
2\pi C_2 & = & 
- \frac{ik}{\eta^3}
 \int du\, e^{ik(u-e\sin u)} \left[ f(u)(1-e \cos u)  + \eta e \sin u  \right].
 \label{eq:hansen_20_parts}
\ee
The integrands in equations \ref{eq:hansen_22_parts} and \ref{eq:hansen_20_parts} are finite near the pole and hence as semi-circle radius around the pole is taken to zero the contribution $C_3=0$. 
The integrand for $X^{2,-2}_k(e)$ contains no $1-e\cos u$ factors in the denominator and does not require integration by parts.

For large $k$, Gaussian integrals are obtained with the dominant contribution near the saddle point. We find that the first terms in a large $k$ expansion are
\be
X^{2,-2}_k(e) & = & e^{-k(y_0-\eta)} \frac{e^2}{4 \sqrt{2\pi} \eta^{9/2} k^{1/2}} \nonumber \\ 
X^{2,0}_k(e) & = & e^{-k(y_0-\eta)} \frac{k^{1/2} }{\sqrt{2\pi} \eta^{3/2}} \nonumber \\ 
X^{2,2}_k(e) & = &e^{-k(y_0-\eta)} \frac{(2 \eta k)^{3/2}}{3\sqrt{\pi} e^2},
\label{eq:hansen_asymp_approx}
\ee
where again these formulas are valid for $k>0$, and $k<0$ values may use the identity $X^{\ell, -m}_{-k} = X^{\ell m}_k$. The exponential factor in Equation \ref{eq:hansen_asymp_approx} appeared in the text in Equation \ref{eq:Klm_exponent}.

\begin{figure*}[htb]
    \epsscale{1.1}
    \plottwo{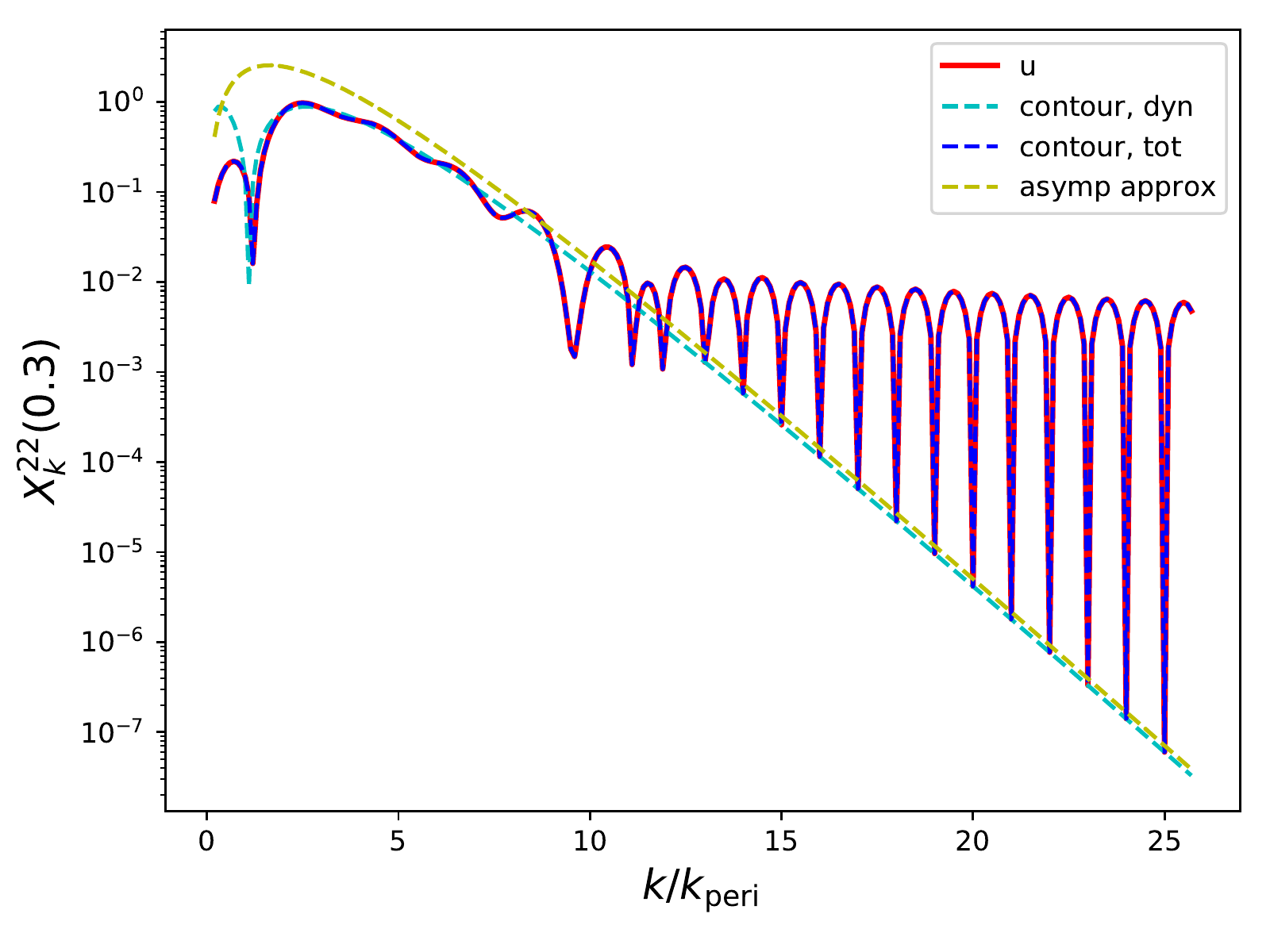}{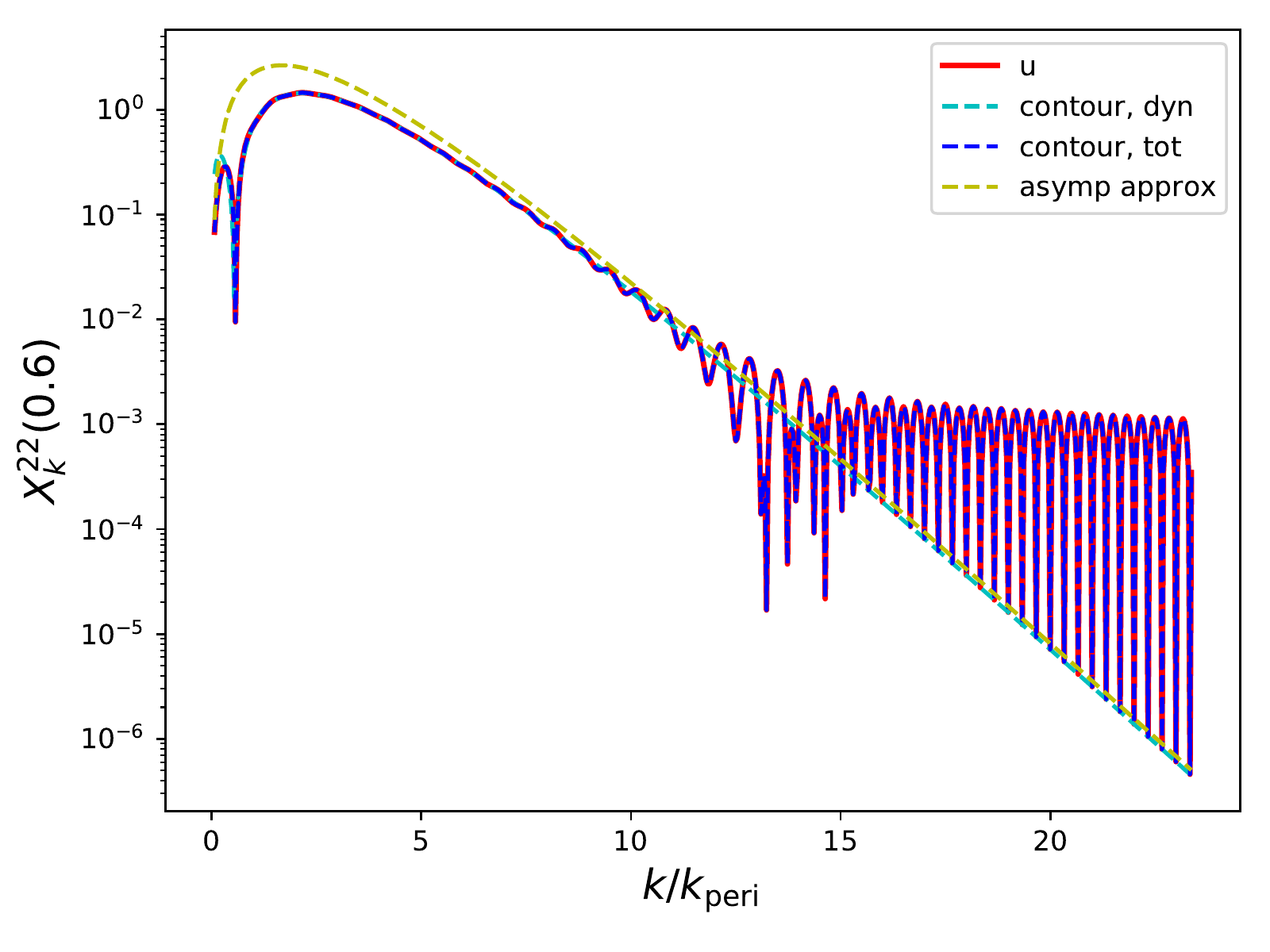} 
    \caption{ Hansen coefficients for $\ell=m=2$ as a function of $k$ for $e=0.3$ (left) and $e=0.6$ (right). The solid red line uses the original integral in Equation \ref{eq:hansen_original}, the dashed blue line is the full contour integration result, while the dashed cyan line is just the dynamical tide piece from contour integration. The dashed yellow line is the asymptotic expansion in Equation \ref{eq:hansen_asymp_approx}. }
    \label{fig:hansen_eq_vs_dyn}
\end{figure*}

Figure \ref{fig:hansen_eq_vs_dyn} compares the original $u$ integration with the contour integral, showing good agreement. For the $e=0.3$ and $e=0.6$ cases shown, the dynamical tide dominates at small $k \la 10k_{\rm peri}$,  beyond which the oscillatory equilibrium tide contribution dominates. The spacing of $k$ values is $0.1$, including integer values where the equilibrium tide contribution is zero. That is where the full curve, when dominated by the equilibrium tide, extends down to the dynamical tide curve. The asymptotic expansion shows agreement for large $k$. Including additional terms in the expansion may improve the accuracy, although convergence is slow. Plots for $m=0$ and $m=-2$ are similar, but their dynamical tide component is smaller relative to the equilibrium tide.

Figure \ref{fig:hansen_eq_vs_dyn} may be compared to Figure 1 in \citet{Wu:18}. In that case, the same integral was performed over a fixed bound orbit, with the x-axis plotted as mode period instead of $k/k_{\rm peri}$. A dynamical tide piece is clearly evident, with amplitude falling steeply as $k$ increases. For sufficiently short mode periods, a transition is found from smooth exponential decrease to a rapidly oscillating behavior, which may be identified as the equilibrium tide contribution discussed here. In their plot, the curves do not always extend down to the dynamical tide curve, likely due to the spacing of the points used. The rapid oscillation is due to the $\sin(\pi k)$ dependence of the equilibrium tide contribution, which oscillates as the mode shifts into and out of resonance with the orbit.

For the mapping simulations described in this paper, the dynamical tide portion of $X^{lm}_k(e)$ is tabulated using numerical integration for $e=0.0001$ to $0.9999$ in steps of $0.0001$. For each $e$, 1000 values of k are tabulated in between $k=1$ and $k=10(1-e)^{-3/2}$. For $k>10(1-e)^{-3/2}$, the asymptotic expansions in Equation \ref{eq:hansen_asymp_approx} are used.

\bibliography{ref.bib}{}
\bibliographystyle{aasjournal}

\end{document}